\documentclass{emulateapj}
\def\lsim{\raise0.3ex\hbox{$<$}\kern-0.75em{\lower0.65ex\hbox{$\sim$}}}
\def\gsim{\raise0.3ex\hbox{$>$}\kern-0.75em{\lower0.65ex\hbox{$\sim$}}}

\begin{document}
\title{Quiescent Cores and the Efficiency of Turbulence-Accelerated, 
Magnetically Regulated Star Formation} 

\author{Fumitaka Nakamura}
\affil{Faculty of Education and Human Sciences, Niigata University,
8050 Ikarashi-2, Niigata 950-2181, Japan; fnakamur@ed.niigata-u.ac.jp}
\and
\author{Zhi-Yun Li}
\affil{Department of Astronomy, University of Virginia, P.O. Box 3818,
Charlottesville, VA 22903; zl4h@virginia.edu}

\begin{abstract}

The efficiency of star formation, defined as the ratio of the stellar 
to total (gas and stellar) mass, is observed to vary from a few 
percent in regions of dispersed star formation to about a 
third in cluster-forming cores. This difference may 
reflect the relative importance of magnetic fields and turbulence 
in controlling star formation. We investigate the interplay between 
supersonic turbulence and magnetic fields using numerical simulations,
in a sheet-like geometry. The geometry allows for an accurate and 
expedient treatment of ambipolar diffusion, a key ingredient for
star formation. We demonstrate that star formation with an efficiency 
of a few percent can occur over several gravitational collapse 
times in moderately magnetically subcritical clouds that are 
supersonically turbulent. In turbulent clouds that are marginally 
magnetically supercritical, the star formation efficiency is higher, 
but can still be consistent with the values inferred for nearby 
embedded clusters. A phenomenological prescription for protostellar 
outflow is included in our model to stop mass accretion after a 
star has obtained a given mass and to disperse away the remaining
core material. Within a reasonable range of strength, the outflow
does not affect the efficiency of star formation much and contributes 
little to turbulence replenishment in subcritical and marginally 
supercritical clouds. If not regulated by magnetic fields at all, 
star formation in a multi-Jeans mass cloud endowed 
with a strong initial turbulence proceeds rapidly, with the majority 
of cloud mass converted into stars in a gravitational collapse 
time. The efficiency is formally higher than the values inferred 
for nearby cluster-forming cores, indicating that magnetic fields 
are dynamically important even for cluster formation. 

In turbulent, magnetically subcritical clouds, the turbulence accelerates 
star formation by reducing the time for dense core formation. The dense 
cores produced are predominantly quiescent, with subsonic internal 
motions. These cores tend to be moderately supercritical, and thus 
remain magnetically supported to a large extent. They contain a small 
fraction of the cloud mass, and have lifetimes long compared with their 
local gravitational collapse time. Some of the cores collapse to form 
stars, while others disperse away without star formation. All these 
factors, as well as core-outflow interaction, contribute to the low 
efficiency of the star formation in these clouds of dispersed star 
formation. 

\end{abstract}

\section{Introduction}

\subsection{Motivation and Previous Work}

Stars are formed out of molecular clouds that are both turbulent and 
magnetized in the present-day Galaxy. The relative importance of 
magnetic fields and turbulence in star formation is a matter of 
debate. One school of thought is that star formation is primarily 
controlled by turbulence (Larson 1981), with magnetic fields playing 
a minor, if any, role (Mac Low \& Klessen 2004). In this picture, 
supersonic turbulence produces dense cores in shocks (e.g., Padoan 
et al. 2001; Gammie et al. 2003; Li et al. 2004), some of which can 
become self-gravitating and collapse to form stars. The shock-produced 
dense cores tend to evolve dynamically (Ballesteros-Paredes, Klessen 
\& Vazquez-Semadeni 2003), unless the postshock region happens to 
contain a mass close to the local thermal Jeans mass. The dynamic 
motions are in contrast with the observation that the starless cores 
of nearby dark clouds tend to be quiescent, with subsonic nonthermal 
velocity dispersions (e.g., Jijina, Myers \& Adams 1999). In the 
best candidates of prestellar collapse (such as L1544), the infall 
speeds derived from line-profile modeling are typically subsonic as 
well (Tafalla et al. 1998; Lee, Myers \& Tafalla 1999). Furthermore, 
a strongly turbulent, weakly magnetized cloud that contains many 
thermal Jeans masses, if not driven constantly on a scale smaller 
than the Jeans length (Klessen, Heitsh \& Mac Low 2000), would collapse 
in a turbulence crossing time, forming stars with a high efficiency 
(e.g., Bate, Bonnell \&  Bromm 2003). The high efficiency may or 
may not be a problem for localized regions of cluster formation 
(Lada \& Lada 2003); it is well above the few percent level that is 
typically observed in 
molecular clouds as a whole (Evans 1999). The problem is particularly 
severe for regions of dispersed low-mass star formation, such as the 
Taurus molecular clouds.  

Inefficient star formation is characteristic of the standard scenario 
of isolated low-mass star formation (Shu, Adams \& Lizano 1987). In 
this picture, a strong, ordered magnetic field is postulated to prevent 
the bulk of the cloud material from prompt collapse. Stars are formed 
in dense pockets of the magnetically supported, subcritical background, 
where the magnetic support has weakened through ambipolar diffusion 
(Nakano 1984). Many authors have done detailed calculations of cloud 
evolution based on this scenario (see Mouschovias \& Ciolek 1999 
and references therein), assuming for numerical convenience that the 
clouds are non-turbulent. Zweibel (2002) and Fatuzzo \& Adams (2002) 
have shown analytically that turbulence can speed up ambipolar 
diffusion in the relatively low density regions where turbulent 
motions are dynamically important. 

Whether the bulk cloud material is indeed magnetically subcritical 
remains to be determined observationally. Available Zeeman measurements  
(Crutcher 1999) indicate that the magnetic flux-to-mass ratios in 
molecular clouds are not far from the critical value ($2\pi G^{1/2}$;
Nakano \& Nakamura 1978) for magnetic support, although the exact 
values are uncertain, because of uncertain projection effects (Shu 
et al. 1999). Shu et al. advanced an argument for the molecular 
clouds in the present-day Galaxy being close to magnetically 
critical: those far from the criticality would have either expanded 
to become diffuse HI clouds or collapsed to become the interiors 
of stars long time ago. In perhaps the best studied starless core,
L1544, the line-of-sight field strength was measured at $\sim 11\ 
\mu$G (Crutcher \& Troland 2000), corresponding to a deprojected 
flux-to-mass ratio of about half the critical value, in agreement
with the prediction of models of core formation driven by ambipolar 
diffusion (Basu \& Mouschovias 1994; Nakamura \& Li 2003). The 
agreement lends support to the picture of low-mass star formation
involving ambipolar diffusion. A potential problem of this picture 
is that it takes an unacceptably 
long time for ambipolar diffusion to produce dense cores directly 
out of the low-density background regions that are well ionized by 
interstellar 
UV photons (McKee 1989; Myers \& Khersonsky 1995). This problem can 
be resolved by strong turbulence, which accelerates the ambipolar 
diffusion-regulated star formation through shock compression. 

Detailed investigations of the combined effects of a strong turbulence 
and magnetic field on star formation in the presence of ambipolar 
diffusion require numerical simulation. In a recent work, we have 
demonstrated, using MHD simulations in a 2D sheet-like geometry, 
that supersonic turbulence and ambipolar diffusion can work together 
to convert a small fraction of the mass of a moderately magnetically 
subcritical cloud into dense collapsed objects over several turbulence 
crossing times (Li \& Nakamura 2004; Paper~I hereafter). The magnetic 
fields prevent the turbulence from compressing most the cloud mass 
into collapse in one crossing time, which would happen if the cloud 
is weakly magnetized. The turbulence, on the other hand, accelerates 
the reduction of magnetic flux in shocked regions. In this paper, we 
will continue our effort in developing this ``turbulence-accelerated, 
magnetically regulated'' scenario of low-mass star formation, which 
appears particularly promising for the relatively inefficient, 
dispersed mode of star formation observed in nearby regions like 
the Taurus molecular clouds. This mode is difficult to accommodate 
in the purely turbulent picture of star formation. 

\subsection{Protostellar Outflows and Stellar Mass Determination}

In Paper I, we took a first step towards determining the efficiency
of star formation, defined as the ratio of the stellar to total 
mass of stars and gas (Lada \& Lada 2003). We  
computed the fraction of cloud mass that has become dense 
(10 times above the average column density) and magnetically 
supercritical. This dense supercritical material can be loosely 
identified with the dense cores of molecular clouds, which are 
known to be intimately tied to low-mass star formation. There is, 
however, evidence that only a fraction of the core material 
eventually ends up inside stars. Onishi et al. (2002), for example, 
derived an average virial mass of 
about 5$M_\odot$ for the H$^{13}$CO$^+$ cores in Taurus, which is an 
order of magnitude higher than the mass of a typical star formed in 
that region (Kenyon \& Hartmann 1995; Palla \& Stahler 2002). It 
is unclear how the 
stellar mass is limited to only a small fraction of the core mass. 
A widely discussed possibility is that the mass 
accretion is stopped by a powerful protostellar outflow (Shu et al.
1987; Matzner \& McKee 2000), particularly 
if the core is surrounded by a magnetically subcritical envelope.
The subcritical
envelope would reduce the mass accretion rate at late times, making 
it easier for the outflow to reverse the infall (Shu et al. 2004). 
The infall-outflow interaction appears to be observed in B5 (Velusamy 
\& Langer 1998), among others. The details of the interaction 
remain to be worked out.

In the absence of a detailed theory of how protostellar mass accretion 
is stopped, we will turn to observation for guidance. In low-mass star 
forming regions such as the Taurus clouds, the typical stellar mass 
is of order $0.5\ M_\odot$. As a rough approximation, we will limit 
the masses of all formed stars to the same value, $M_*$. As soon as 
the stellar mass reaches $M_*$, the accretion is terminated suddenly. 
At the same time, we eject the dense material surrounding the formed 
star in an outflow, with a speed set by the requirement that the 
outflowing gas carry away a certain amount of (linear) momentum 
(see also Allen \& Shu 2000). We 
envision the outflowing gas as the core material swept up by a fast 
protostellar wind emanating from close to the central star in a 
momentum-conserving fashion. The wind momentum is given by  
\begin{equation}
P_{\rm wind}=M_{\rm wind} V_{\rm wind}=100\; f\; M_*\; (M_\odot\; 
{\rm km/s}), 
\label{windmomentum}  
\end{equation}
where the dimensionless parameter $f$ is the product of the wind speed 
$V_{\rm wind}$ in units of 100 km/s and the ratio of the mass ejected
in the wind $M_{\rm wind}$ to the stellar mass $M_*$. 
The protostellar outflows included in our 
simulation provide a potential means to replenish 
the supersonic turbulence, which decays quickly despite the presence 
of a dynamically important magnetic field (Stone et al. 1998; Mac Low
et al. 1998; Padoan \& Nordlund 1999). 
They may also directly induce star formation.  

As in Paper I, we will adopt a sheet-like cloud geometry, where 
ambipolar diffusion can be computed accurately and expeditiously. 
Our goal is to investigate the interplay between magnetic fields
and supersonic turbulence, taking into account the effects of 
the stars formed in the cloud and their protostellar outflows. 
Specifically, we want to address two fundamental issues: the 
efficiency of star formation and the formation of quiescent cores
out of a turbulent background medium. The remainder of the paper is 
organized as follows. In \S~2, we describe our mathematical formulation 
of the problem. Numerical results are described and interpreted in \S~3 
through \S~5, and their implications are discussed in \S~6. The last 
section, \S~7, contains a brief summary.  

\section{Formulation of the Problem}

\subsection{Governing Equations}

We consider clouds threaded by well ordered magnetic fields, and adopt 
the standard thin-sheet approximation, taking advantage of the tendency 
for strongly magnetized clouds to settle along the field lines into 
a flattened configuration in a dynamical time (e.g., Fiedler \& 
Mouschovias 1993; Ostriker, Gammie \& Stone 1999). We restrict the 
gas motions to the plane of matter distribution, taken to be the 
$x$-$y$ plane of a Cartesian system $(x,y,z)$. Possible coupling 
between the motions in the plane and in the vertical direction will 
be addressed in future 3D simulations (see Krasnopolsky \& Gammie
2005). The 2D cloud evolution is 
governed by a set of vertically integrated MHD equations. These 
equations are the same as those used in our previous investigations 
(Li \& Nakamura 2002; Nakamura \& Li 2003; Paper~I). They are 
described below for reference. 

The vertically integrated equation for mass conservation is 
\begin{equation}
\frac{\partial \Sigma }{\partial t} +
\nabla \cdot \left(\Sigma \mbox{\boldmath$V$} \right) = 0 \; ,
\label{eq:basic1}
\end{equation}
where $\mbox{\boldmath$V$} = \left(V_x, V_y \right)$ is the velocity of
the bulk, neutral cloud material in the plane of mass distribution. 
The momentum equation is  
\begin{equation}
\Sigma \frac{\partial \mbox{\boldmath$V$}}{\partial t} +
 \left(\Sigma \mbox{\boldmath$V$}\cdot \nabla\right) 
  \mbox{\boldmath$V$} 
   + \nabla P + H \nabla \left(\frac{B_z ^2}{4\pi}\right)
   - \frac{B_z \mbox{\boldmath$B$}}{2\pi} - \Sigma 
   \mbox{\boldmath$g$}= 0 \; ,
 \label{eq:basic2}
\end{equation}
where $\mbox{\boldmath$B$} = \left(B_x, B_y \right)$ and $B_z$ are 
the horizontal and vertical components of the magnetic field, and
$\mbox{\boldmath$g$} = \left(g_x, g_y \right)$ is the horizontal
component of the gravitational acceleration. 
The quantity $P$ is the vertically integrated pressure, and $H$ 
the half-thickness of the sheet. The latter is related to
the column and mass densities $\Sigma$ and $\rho$ through 
\begin{equation}
H = \Sigma /(2\rho).
\label{eq:thickness}
\end{equation}
The equation governing the evolution of the vertical magnetic field 
component is
\begin{equation}
 \frac{\partial B_z}{\partial t} +
  \nabla \cdot \left(B_z \mbox{\boldmath$V_{\rm B}$} \right) = 0 \; ,
 \label{eq:basic3}
\end{equation}
where $\mbox{\boldmath$V_{\rm B}$}=(V_{{\rm B},x},V_{{\rm B},y})$ is the
velocity vector of magnetic field lines in the plane.

The governing equations (\ref{eq:basic1}), (\ref{eq:basic2}), and 
(\ref{eq:basic3}) are supplemented by the usual isothermal equation 
of state, which in a vertically integrated form becomes   
\begin{equation}
P = c_s^2 \Sigma ,
\label{eq:EOS}
\end{equation}
where $c_s$ is the isothermal sound speed. The isothermal condition 
(\ref{eq:EOS}) breaks down in low density regions where heating by 
UV radiation is important and in high density regions where
thermal radiation becomes trapped; the deviation from isothermality 
is ignored at low densities since the thermal pressure is dominated 
by the turbulent pressure, and at high densities because the thermal
pressure plays a minor role in the densest regions that are collapsing
dynamically. 

The velocity of magnetic field lines in the induction 
equation~(\ref{eq:basic3}) can be computed from 
\begin{equation}
  \mbox{\boldmath$V_{\rm B}$} - \mbox{\boldmath$V$}
   = {t_c \over \Sigma}  \left[\frac{B_z \mbox{\boldmath$B$}}
	{2\pi}- H \nabla \left(\frac{B_z^2}{4\pi}\right) \right] ,
 \label{eq:basic4}
\end{equation}
where $t_c$ is the coupling time between the magnetic field and 
neutral matter. Realistic computations of $t_c$ are complicated
by dust grains, whose size distributions are uncertain in dense
clouds (e.g., Nishi, Nakano \& Umebayashi 1991). For our purposes, 
we adopt the expression 
\begin{equation}
t_c = \frac{1.4}{C\rho^{1/2}} \; ,
 \label{eq:basic5}
\end{equation}
which is valid in the simplest case where the coupling is provided by 
ions that are well tied to the field lines and the ion density $\rho_i 
\propto \rho ^{1/2}$. For the coefficient $C$, we adopt the fiducial 
value $1.05\times 10^{-2}$~cm$^{3/2}$g$^{-1/2}$s$^{-1}$ (Shu 1991). 
The factor 1.4 in the above expression accounts for the neglect of 
ion-helium collision, whose cross section is small compared to that 
of ion-hydrogen collision (Nakano 1984; Mouschovias \& Morton 1991). 
In relatively diffuse regions where the visual extinction is less than 
a few magnitudes, photoionization becomes important, which can enhance 
the magnetic coupling (McKee 1989). An enhanced 
coupling at densities below the average value adopted in our 
calculations does not change the results qualitatively.

The mass density that appears in equations (\ref{eq:thickness}) and 
(\ref{eq:basic5}) can be determined from the condition of approximate 
force balance in the vertical direction (Fiedler \& Mouschovias 1993),
which yields  
\begin{equation}
\rho = \frac{\pi G\Sigma ^2}{2 c_s^2}\left(1+\frac{B_x^2 + B_y^2}
{4\pi^2 G \Sigma^2}\right) + \frac{P_e}{c_s^2} .
 \label{eq:volumndensity}
\end{equation}
The two terms in the brackets represent, respectively, the gravitational 
compression and magnetic squeezing of the cloud material. The quantity 
$P_e$ is the ambient pressure that helps confine the sheet, especially 
in low column density regions where gravitational compression is 
relatively weak.

\subsection{Initial and Boundary Conditions}
\label{ibcs}

The equations (\ref{eq:basic1})-(\ref{eq:volumndensity}) governing the 
cloud evolution are solved numerically. For 
initial conditions, we choose a uniform distribution of mass in 
the $x$-$y$ plane, threaded by a magnetic field of constant strength 
$B_0$ in the $z$ direction. In the absence of magnetic field, the 
cloud would be gravitationally unstable to perturbations of wavelength 
greater than the Jeans length $L_J=c_s^2/(G\Sigma_0)$ 
(Larson 1985). Here, $\Sigma_0$ is the column density normal to 
the sheet. From the Jeans length $L_J$ and sound speed $c_s$, we can 
define a time scale for gravitational collapse $t_g=L_J/c_s=c_s
/(G\Sigma_0)$ (e.g., Ostriker et al. 1999). The Jeans length $L_J$ 
and collapse 
time $t_g$ provide two basic scales for our problem. We adopt a square 
computation box of length $L \gg L_J$ on each side, with periodic 
conditions imposed at the boundaries. The region under 
consideration thus contains a large number ($\sim [L/L_J]^2$) 
of thermal Jeans masses. Individual regions of size comparable to 
$L_J$ are expected to collapse to form stars or stellar 
systems on a time scale comparable to $t_g$, unless other agents 
intervene.

The gravitational instability is 
suppressed if the field strength $B_0$ is greater than a critical 
value $2\pi G^{1/2}\Sigma_0$ (so that the cloud is magnetically
subcritical) in the ideal MHD limit (Nakano \& Nakamura 1978). In 
a typical molecular cloud that is lightly ionized, the instability 
can still grow, 
albeit on the longer time scale of ambipolar diffusion (Langer 1978). 
We have previously followed the nonlinear evolution of gravitational 
instability in weakly ionized, magnetically subcritical clouds, and 
found that ambipolar diffusion-driven fragmentation can readily lead 
to binary and multiple star formation in quiescent cloud cores in 
the presence of small perturbations (Nakamura \& Li 
2002, 2003; Li \& Nakamura 2002; see also Indebetouw \& Zweibel 2000
and Basu \& Ciolek 2004). A strong turbulence is expected to have
a larger effect on cloud fragmentation. Some aspects of star 
formation in turbulent magnetized clouds were examined in Paper~I. 
They are investigated in greater depth in the present paper. 

To mimic the turbulent motions observed in molecular clouds, we 
introduce into the initially uniform cloud a supersonic velocity 
field at the beginning of the simulation. Following the usual  
practice (e.g., Ostriker et al. 1999; Klessen et al. 
2000), we prescribe the turbulent velocity field in Fourier space, 
with a power-spectrum $v_k^2\propto k^{-n}$. Unless noted otherwise,
we will choose a power index of $n=3$, which is compatible with 
the Larson's (1981) size-dispersion relation in two spatial 
dimensions (Gammie 2003, priv. comm.). Numerically, we consider
only discrete values $k_i=2\pi i/L$ ($i=-N/2$, ..., $N/2$, with 
$N$ being the number of grid points in each direction) for 
the two wavenumber components, $k_x$ and $k_y$. For each pair of 
$k_x$ and $k_y$, we randomly select an amplitude from a Gaussian 
distribution consistent with the power spectrum for the total 
wavenumber $k=\sqrt{k_x^2+k_y^2}$ of that pair and a phase between 
0 and $2\pi$. The resulting field is then transformed into the 
physical space to obtain the distribution 
of a velocity component at each grid point. The distributions of 
$V_x$ and $V_y$ are generated independently, and the final velocity 
field is scaled so that the root mean square (rms) Mach number of
the flow ${\cal M}$ has a prescribed value. The velocity field so 
specified tends to have a strong compressive component. For 
comparison, we also investigate 
cases with incompressible turbulent velocity fields, generated by 
imposing the condition that the wavenumber vector be perpendicular 
to the direction of wave propagation (Ostriker et al. 1999).   

\subsection{Dimensional Scalings}
\label{units}

The actual numerical computations are carried out using non-dimensional
quantities. The dimensional units we adopt are $c_s=1.88\times 10^4 
T_{10}^{1/2}$~cm~s$^{-1}$ for speed (where $T_{10}$ is the temperature 
in units of 10~K), $\Sigma_0=4.68\times 10^{-3} A_V$~g~cm$^{-2}$ for 
surface density (where $A_V$ is the visual extinction vertically 
through the sheet for standard grain properties), and $B_0=7.59 A_V 
\Gamma_0$~$\mu$G for magnetic field strength (where $\Gamma_0$ is the 
flux-to-mass ratio in units of the critical value). The units for 
length, time, and mass are, respectively, the Jeans length $L_J\equiv 
c_s^2/(G\Sigma_0)=0.37\ T_{10}/A_V$~pc, gravitational collapse time 
$t_g\equiv c_s/(G\Sigma_0)=1.90\times 10^6 T_{10}^{1/2}/A_V$~yr, and 
Jeans mass $M_J\equiv \Sigma_0 L_J^2 =3.02\ T_{10}^2/A_V$~$M_\odot$. 
We normalize the external pressure $P_e$
by the ``gravitational pressure'' 
$2\pi G \Sigma_0^2$, and set its dimensionless value to $0.025$ for 
all models. As long as the external pressure is significantly lower
than the ``gravitational pressure'', its exact value does not affect 
the cloud evolution much. 

\subsection{MHD Code, Outflow Implementation and Lagrangian Particles}
\label{code}

We carry out numerical calculations using an MHD code that treats
self-consistently the evolution of magnetic field in a sheet-like 
geometry, including ambipolar diffusion. The code is described in 
Li \& Nakamura (2002), modified here to take into account the periodic 
boundary conditions imposed in this problem. It was based on the code 
originally developed in Nakamura \& Hanawa (1997). Briefly, we solve the 
hydrodynamic part using Roe's TVD (Total Variation Diminishing) 
method given in Hirsch et al. (1990). The vertical component of 
magnetic field is evolved 
in a way similar to the column density, since both of them satisfy the 
same form of continuity equation (see equations [\ref{eq:basic1}] and 
[\ref{eq:basic3}]). We define a magnetic potential outside the sheet, 
which is used to determine the magnetic forces on the sheet, in a 
manner similar to the determination of gravitational forces from the 
gravitational potential. Both the magnetic and gravitational potentials 
are solved using a convolution method based on FFT, as described in 
Hockney et al. (1981). The computational timestep is chosen small enough
to satisfy both the usual Courant condition and the condition for
treating ambipolar diffusion stably in an explicit code such as ours. 
We have verified that in the ideal MHD limit, the numerical diffusion 
of magnetic flux in our code is negligible, with the flux-to-mass 
ratio conserved within the machine roundoff error. 

We fix the size of our simulation box to 10 times the Jeans length, 
i.e., $L=10\ L_J=3.7\ T_{10}/A_V$~pc. There are, therefore, $10^2$ 
thermal Jeans masses (or $302\ T_{10}^2
/A_V$~$M_\odot$) in our (two dimensional) computational domain. 
A $512\times 512$ grid is used, with a 
cell size of $\Delta L=7.2\times 10^{-3}\ T_{10}/A_V$~pc or about 
$1500\ T_{10}/A_V$~AU. Obviously, the grid is too coarse to resolve
the flow pattern close to a forming star. Nevertheless, once dynamical
collapse is initiated in a dense region, the mass accumulated in the 
highest density cell should provide a fair estimate of the stellar 
mass. As mentioned in the introduction, the stellar mass is probably
limited by protostellar outflows, although the details of the 
infall-outflow interaction remain uncertain. 
For simplicity, we implement into the MHD code the following procedure
for stellar mass determination: as soon as the column density in a 
cell reaches a threshold $\Sigma_{\rm max}$, we replace it with a 
much lower floor value $\Sigma_{\rm min}$. The mass extracted, $M_*
=(\Sigma_{\rm max}-\Sigma_{\rm min}) (\Delta L)^2$, is placed in a 
Lagrangian particle, which represents a formed star. Our canonical 
choices are $\Sigma_{\rm max}=460\Sigma_0$ and $\Sigma_{\rm min}
=30\Sigma_0$, which yield a 
stellar mass $M_*= 0.5\ T_{10}^2/A_V$~$M_\odot$. We keep the magnetic 
flux in the cell unchanged, to reflect the fact that stars are born
with a flux-to-mass ratio much smaller than the critical value. 
   
At the same time that the stellar mass is extracted from a cell, we set 
the mass $M_{\rm flow}$ within a region of radius $5 \Delta L =3.6 
\times 10^{-2}\ T_{10}/A_V$~pc (centered on the cell) into a radial 
motion (in the plane of mass distribution), with a speed  
\begin{equation}
V_{\rm flow}=P_{\rm wind}/M_{\rm flow}=100\; f\; \left({M_*\over
M_{\rm flow}}\right)\;\; {\rm km/s},
\label{flowspeed}
\end{equation}
where equation~(\ref{windmomentum}) for protostellar wind momentum has 
been used. The outflow parameter $f$ is the product of the wind speed 
in units of 100 km/s and the ratio of the mass ejected in the wind
to the stellar mass. For optically 
revealed classical T Tauri stars, the wind speed is typically a few
hundred km/s, and $\sim 10\%$ of the mass accreted through the disk 
is ejected in the wind (Calvet 1997). If the wind is launched 
magnetocentrifugally from a circumstellar disk (e.g., Konigl \& 
Pudritz 2000), the wind speed is expected to be somewhat lower 
during the embedded, protostellar accretion phase than during the 
revealed phase, because of a lower stellar mass (and thus slower
disk rotation) and a higher mass loading. If protostellar winds 
carry away a similar fraction of the accreted mass as the T Tauri
winds, then $f\approx 0.1$, for a reasonable wind speed of 100 km/s. 
This is the canonical value that we will adopt. A complication is
that most of the wind momentum may escape in a direction perpendicular 
to the mass distribution in 3D, which would lower the value of $f$. 
To allow for this possibility, we also consider cases with a much 
reduced $f=0.01$. 

In our model, the termination of mass accretion and turn-on of outflow 
are assumed to be sudden. These can be justified on the ground that 
the bulk of mass accretion and outflow occur during the Class 0 
phase of low-mass star formation, which lasts for only a few times 
$10^4$ years (Andre et al. 2000), much less than the collapse time 
$t_g=1.90\times 10^6 T_{10}^{1/2}/A_V$~yr at the average cloud 
density. 

Once created, the (stellar) Lagrangian particles are evolved in the 
collective gravitational potential of both the cloud material and 
the stars. To avoid singularities, we soften the gravitational 
potential of each star within a radius $3 \Delta L=2.2\times 
10^{-3}\ T_{10}/A_V$~pc of the star. Although the choice of the 
radius is somewhat arbitrary, the gravitational softening may mimic 
the effects of a stellar wind, which may prevent a formed star 
from accreting additional ambient material as it moves around in 
the potential well of the cloud.  

%
%

\section{Efficiency of Star Formation in Turbulent Magnetized Clouds} 

\subsection{Standard Simulation}

A number of parameters are needed to specify the turbulence, magnetic 
field and protostellar outflow in our problem. Some of these will be 
explored in subsequent subsections. Here, we focus on a standard 
simulation with a set of standard parameters to illustrate the basic 
results. It will also serve as a standard against which other 
simulations are compared. Since one of our emphases is on ambipolar 
diffusion, which is expected to play a more important role in 
magnetically subcritical clouds than in supercritical clouds, we 
choose a ``standard'' dimensionless flux-to-mass ratio $\Gamma_0=1.2$. 
The cloud is thus moderately subcritical. At the beginning of 
computation $t=0$, we stir the cloud with a random velocity field 
of power index $n=3$ (so that the turbulent energy is dominated by 
large-scale motions) and rms Mach number ${\cal M}=10$, generated 
using the procedure outlined in \S~\ref{ibcs}. Unless noted otherwise, 
this ``standard'' random realization of turbulent velocity field is 
applied to other simulations. The star created in the collapse of a 
dense core is handled by a Lagrangian particle according to the 
prescription outlined in \S~\ref{code}. The remaining core material 
is ejected in an outflow, with the outflow parameter $f$ set to the 
canonical value $0.1$. The parameters of the standard simulation
and its variants are listed in Table~1.

We have followed the cloud evolution in the standard simulation 
(Model~S1 in Table~1) 
well beyond the gravitational collapse time $t_g=1.90\ T_{10}^{1/2}
/A_V$~Myr of the initial 
cloud. A movie of the evolution can be obtained from the authors. 
The main results are shown in Figs.~1-3. In Fig.~1, we plot the 
most important quantity that we are after, the star formation 
efficiency (SFE hereafter), as a function of time. The first star 
forms around $t=0.65\ t_g=1.24\ T_{10}^{1/2}/A_V$~Myr. Thereafter, 
the number of stars increases slowly but steadily. By the time $t
=4\ t_g=7.60\ T_{10}^{1/2}/A_V$~Myr, 32 stars of $0.5 T_{10}^2/A_V\ 
M_\odot$ each have formed, yielding an accumulative SFE of $5.3\%$. 

The low efficiency of star formation is the hallmark of the standard 
simulation. The efficiency is low because of the strong magnetic 
field imposed on the cloud initially. The field strength is such 
that the majority of the cloud mass remains magnetically subcritical 
even in the presence of a supersonic turbulence. Pockets of 
supercritical material that are capable of forming stars are created 
through ambipolar diffusion, which is a relatively slow process. The
ambipolar diffusion is governed by equation~(\ref{eq:basic4}), which 
can be cast into a more transparent form
\begin{equation}
\frac{\mbox{\boldmath$V_{\rm B}$} - \mbox{\boldmath$V$}}{c_s}
   = 0.17 {\Gamma^2 \over D^{1/2}} \left(\frac{\mbox{\boldmath$B$}}
	{B_z}- \frac{H {\nabla B_z}} {B_z}\right),
 \label{ad}
\end{equation}
where the auxiliary quantity 
\begin{equation}
D=1+{\Gamma^2 (B_x^2+B_y^2)\over B_z^2}
\label{Afactor}
\end{equation}
is the column density enhancement factor due to magnetic compression 
in the vertical direction (with the usually small external pressure 
term in equation~[\ref{eq:volumndensity}] ignored). For a cloud near 
magnetic criticality $\Gamma\sim 1$, the drift speed of magnetic field 
lines relative to the bulk neutral material, $\mbox{\boldmath$V_{\rm B}$} 
- \mbox{\boldmath$V$}$, is much smaller than the sound speed $c_s$ 
in regions of mild field variation (where the field lines are nearly 
vertical, i.e., $B_x^2+B_y^2\ll B_z^2$, and $B_z$ varies gradually 
in space). The diffusion is faster in shocks where $B_z$ changes 
over a shorter distance and the field lines are bent more strongly 
away from the vertical direction. The enhanced diffusion in shocks 
leads to accelerated production of supercritical material in localized 
regions of the turbulent, magnetically subcritical clouds.

The magnetically supercritical regions created through ambipolar diffusion 
are shown in Fig.~2. The figure displays the snapshots of the standard 
cloud at six representative times $t=0.5, 1.0, 1.5, 2.0, 3.0$, and $4.0\ 
t_g$, with white contours separating the ``barren'' subcritical material 
that cannot form stars from the ``fertile'' supercritical material that 
has the potential for star formation. At early times, supercritical 
material is created in numerous small pockets, as a result of the 
power spectrum of the initially imposed turbulent velocity field. The 
velocity distribution is such that regions of small characteristic 
lengthscale collide first, creating a large number of shocklets (see 
also Gammie et al. 2003), where ambipolar diffusion is accelerated. 
The amount of mass involved in each shocklet is small, less than the 
``effective'' Jeans mass, which is increased over the purely thermal 
value by both the magnetic pressure and partial cancellation of the 
gravitational force by the magnetic tension force (Nakamura \& Hanawa  
1997; Shu \& Li 1997). These supercritical regions do not collapse 
promptly. They are swept up at later times by the larger scale, 
converging turbulent motions into a network of more massive filaments. 
The supercritical filaments tend to be denser than the subcritical 
background because their self-gravity is stronger than the magnetic 
tension force, especially at late times. The cancellation of the 
tension force by self-gravity ``softens'' the supercritical regions, 
making them easier to compress by turbulence and harder to rebound 
after compression. 

The mass fraction of the supercritical material is plotted in Fig.~3 
as a function of time. It increases sharply initially, as a result of 
the large gradients in the initial turbulent velocity field, which 
create strong shocklets that accelerate ambipolar diffusion. The 
second increase occurs around one turbulence crossing time, defined 
here as the time for the large-scale turbulence to cross half of 
the simulation box,  
\begin{equation}
t_{\rm x}={L\over 2{\cal M} c_s}.
\label{crossing}
\end{equation}
For the standard simulation, the crossing time $t_{\rm x}=0.5\ t_g$. 
The increase is probably caused by the large-scale shock compression. 
Thereafter, the mass fraction stays more or less constant (roughly a 
third) over a period of several (average) gravitational collapse 
times. This value provides an upper limit to the efficiency of star 
formation. 

Most of the supercritical filaments remain magnetically supported to 
a large extent. They last for a time long compared with the local 
dynamical collapse time. The longevity is unique to the shock-produced 
filaments that occur in magnetically subcritical clouds through 
ambipolar diffusion. Similar filaments created in weakly magnetized, 
supercritical clouds are short-lived; they fragment and collapse 
promptly in a 
dynamical time. On the other hand, in the absence of ambipolar diffusion 
(i.e., ideal MHD limit), converging turbulent flows can still produce 
dense filaments in subcritical clouds. The filaments remain subcritical, 
however. They quickly disperse away once the external confinement 
disappears (Vazquez-Semedani et al. 2005). Evidently, the ambipolar 
diffusion in our standard simulation has weakened the magnetic field 
inside the filaments enough for them to ``stick'' together after 
compression, but not so much that their self-gravity would completely 
overwhelm the magnetic forces and collapse the filaments dynamically. 
The magnetic support of supercritical filaments, which contain a 
minor fraction of the cloud mass to begin with, further reduces the 
rate and efficiency of star formation. 

Stars are formed in dense cores of supercritical filaments. These cores 
will be discussed in depth in the next section, \S~4. Here we merely 
point out that only a small fraction of the supercritical material 
resides in the densest regions that are most intimately linked to star 
formation. Fig.~3 shows that the mass fraction is about $10\%$ at 
column densities above 10 times the average value, typical of the 
low-mass dense cores identified in nearby dark clouds. The dearth of 
dense gas further limits the star formation efficiency. 

As we will see in \S~4, some of the dense cores expand and disperse
away without star formation. For those that do collapse to form 
stars, the local efficiency is limited by outflows. In our simulation, 
mass accretion from the core is cut off once the stellar mass reaches 
a pre-specified value (typical of the low-mass stars formed in nearby 
dark clouds). The remaining core material is blown away in an outflow. 
Some of the cavities evacuated by outflows are evident in Fig.~2. On 
a global scale, the outflows can in principle stimulate further star 
formation by compressing supercritical regions into collapse, and by 
speeding up ambipolar diffusion in the material that they sweep up. 
They do not, however, fundamentally change the inefficient nature of 
star formation in magnetically subcritical clouds such as the one in 
our standard simulation. 

\subsection{Varying Outflow Strength of Standard Simulation}

The amount of protostellar outflow momentum that is coupled to the 
ambient medium is uncertain. In particular, there is the possibility 
that most of the outflow momentum escapes perpendicular to the plane 
of mass distribution. To explore this possibility, we rerun the 
standard simulation with the outflow parameter $f$ reduced by a 
factor of 10, to 0.01 (Model~S2 of Table~1). 
The results are shown in Fig.~4, where the snapshots 
of the cloud at six representative times $t=0.5, 1.0, 1.5, 2.0, 3.0$ 
and $4.0\ t_g$ are displayed, as in Fig.~2 for the standard simulation. 
Comparison of the two figures shows that the cloud morphologies are 
rather similar, especially at early times, despite the factor-of-ten 
difference in outflow strength. The main difference lies in outflow 
cavities which, as expected, are less prominent in the weaker 
outflow case. 

One may naively expect the stronger outflows in the $f=0.1$ case to 
stimulate more star formation. This turns out not to be the case. 
The number of stars formed near the end of the simulation at $t=4\ 
t_g$ is lower in the $f=0.1$ case (32) than in the $f=0.01$ case 
(40) (see also Fig.~6 below). The reason, we believe, is that a 
stronger outflow tends to disperse the supercritical material around 
a formed star over a larger region, making it more difficult to 
re-condense. In any case, the difference in the efficiency of star 
formation is modest (only 25\%), which leads us to believe 
that the SFE is relatively insensitive to the strength of protostellar
outflow. 

\subsection{Varying Turbulent Velocity Field of Standard Simulation}

The kinetic energy of the turbulence in molecular clouds is thought 
to be dominated by large-scale motions (Larson 1981). The exact 
velocity field is unknown, however. It is usually modeled through 
random realization of a power spectrum specified in Fourier space, 
as described in \S~\ref{ibcs}. To check the dependence of our results 
on the prescribed turbulent velocity field, we carried out a set of 
6 simulations, keeping the rms Mach number fixed at ${\cal M}=10$. 
The model parameters are listed in Table~1. Simulations R1 and R2 
are the same 
as the standard simulation except that the turbulent velocity field 
of power index $n=3$ is realized in a different way. Simulations I1 
and I2 also have power spectra of $n=3$ but the velocity fields are 
chosen to be incompressible. Simulations N1 and N2 have random 
velocity fields of power spectra with $n=5$ and $1$, respectively. 
The results are shown in Fig.~5. 

All four different random realizations of turbulent velocity field 
of the same $n=3$ power spectrum yield a SFE close to that of the 
standard simulation. The SFE of the $n=5$ case, where the power is 
more dominated by the large scales, is hardly distinguishable from 
those of the $n=3$ cases. The $n=1$ case, on the other hand, is quite 
distinct from all other cases. Its star formation did not start until 
about $t=3.5\ t_g$, after most of the initial turbulence has already 
decayed away. The lower efficiency in this case is probably due to 
a larger fraction of the turbulent energy residing on small scales, 
which dissipates quickly without compressing enough mass to enable
gravitational collapse. Based on this set of simulations, we conclude 
that the low SFE we obtained in the standard simulation is robust. 
It is, we believe, a generic feature of star formation in turbulent, 
moderately subcritical clouds. We now turn our attention to 
non-magnetic and marginally magnetically supercritical clouds, which 
behave differently.

\subsection{Marginally Magnetically Supercritical and Non-magnetic Clouds}
\label{supnon}

Weaker magnetic fields lead to higher rates of star formation. We 
demonstrate this trend with two examples. Both have the same initial 
mass distribution and turbulent velocity field as in the standard 
simulation. The only difference is that their flux-to-mass ratio 
$\Gamma_0=0.8$ (Model~U1) and 0 (Model~O1), respectively, instead 
of 1.2. The results are shown in Figs.~6 and 7. We consider these 
two cases in turn.

The marginally magnetically supercritical cloud ($\Gamma_0=0.8$) forms
stars continuously over a time much longer than the turbulence crossing
time ($t_{\rm x}=0.5\ t_g$), with a rate that decreases with time 
(Fig.~6). 
The decrease is due, at least in part, to the fact that, as more (weakly 
magnetized) stars form, the remaining gas becomes more magnetized. 
The average flux-to-mass ratio eventually exceeds the critical value 
and the cloud becomes subcritical as a whole. Another reason may be 
the decay of turbulence, which is at a relatively low level despite 
the momentum input from protostellar outflows (see \S~\ref{decay} below). 
Near the end of the 
simulation at $t=3.5\ t_g$, 162 stars are produced, corresponding to 
a SFE of about $27\%$. The efficiency is in the range inferred for 
nearby embedded clusters (from $\sim 10$ to $\sim 30\%$; Lada \& Lada 
2003). This similarity in SFE indicates to us that cluster-forming 
regions, 
although supercritical, may still be substantially magnetized. For 
comparison, we plot in the same figure the SFE for the non-magnetic 
cloud ($\Gamma_0=0$). In this case, stars are formed much more quickly. 
More than half of the cloud material is converted into stars by the 
end of the simulation at $t=1.5~t_g$. The high efficiency is obtained 
despite the presence of a vigorous outflow from each of the formed 
stars. It is formally above the values given in Lada \& Lada. 

As the outflow parameter decreases from the standard value $f=0.1$ 
to 0.01, the efficiency of star formation becomes somewhat higher. 
Fig.~6 shows that, in the marginally supercritical ($\Gamma=0.8$) 
cloud, about 10\% more stars are formed by $t=3\ t_g$ in the $f=0.01$ 
case than in the $f=0.1$ case. The fractional increase is larger 
in the non-magnetic ($\Gamma_0=0$) cloud. Its SFE in the weaker 
outflow case is even more inconsistent with the observationally 
inferred values (Lada \& Lada 2003). 

Besides the rate and efficiency of star formation, the non-magnetic 
and marginally supercritical clouds differ in another important 
aspect: cloud mass distribution. The difference is illustrated in 
Fig.~7, where the snapshots of the non-magnetic case at $t=0.5, 1.0$, 
and $1.5\; t_g$ are shown in the first three panels and those of the 
magnetized case at $t=1.0, 2.0,$ and $3.0\; t_g$ in the last three. 
The contrast in cloud appearance is striking. Most of the gas in the 
non-magnetic cloud is concentrated in very thin dense filaments, 
which are clothed by little moderately dense envelopes. These 
essentially bare filaments are surrounded by large voids, similar 
to those commonly observed in pressureless cosmological simulations. 
In contrast, the filaments in the magnetized cloud are embedded 
within moderately dense gas, particularly at late times. A third
difference is that the flow speed is substantially higher in the
non-magnetic cloud than in the magnetized cloud. This difference
will be discussed further in \S~\ref{decay} in connection with 
turbulence decay. 

\subsection{Effects of Turbulent Mach Number} 

The level of turbulence varies on different scales of molecular clouds 
and from place to place. Regions of stronger turbulence are expected
to produce stars more vigorously. This trend is demonstrated in Fig.~8,
where the SFEs of the standard simulation (with a turbulent Mach number 
${\cal M}=10$) and its two variants (Models~S3 and S4 in Table~1) of 
weaker turbulence (${\cal M}=3$ and $1$, respectively) are plotted. 
As the initial turbulent Mach number decreases, star formation starts 
at a later 
time, with a longer quiescent period of little or no star formation. 
In the case of ${\cal M}=3$, the first star forms around $t=1.3\ t_g$ 
(or $\sim 2.5 T_{10}^{1/2}/A_V$~Myr). It is followed by another non-star 
forming period of similar duration. After about $t=2.7\ t_g$, the rate 
of star formation becomes comparable to, or even higher than the 
initially more turbulent standard case. The late increase in star 
formation activities is driven, we believe, by ``spontaneous'' 
ambipolar diffusion (as opposed to the ``forced'' ambipolar diffusion 
in strong shocks), which operates at all times, particularly in higher 
density regions. The ``spontaneous'' ambipolar diffusion increases 
the mass fraction of supercritical material gradually after 
the initial strong turbulent compression. Even in this gentler mode 
of ambipolar diffusion the turbulence plays a crucial role, by 
producing the density inhomogeneities needed for efficient diffusion. 
Once stars form, their outflows can further enhance the SFE by 
compressing nearby supercritical regions (formed through ambipolar 
diffusion) into collapse.
In this interpretation, the longest quiescent period, observed in the 
least turbulent ${\cal M}=1$ case, is related to the fact that its
density peaks have the lowest values. Comparison of these three cases 
clearly demonstrates that in magnetically subcritical clouds star 
formation is accelerated by turbulence. 

Star formation in magnetically {\it critical} clouds is accelerated by 
turbulence as well. The speedup of star formation with increasing
turbulent Mach number is shown in Fig.~8, where the SFEs of a cloud 
identical to that in the standard simulation except for $\Gamma_0
=1.0$ (instead of $1.2$) are plotted for three values of Mach number 
${\cal M}=10$, 3 and 1 (corresponding to Models~C1, C2, and C3 in Table~1,
respectively). Note that, for a given ${\cal M}$, the rate of star 
formation is higher for the critical ($\Gamma_0=1.0$) cloud than for 
the standard (subcritical, $\Gamma_0=1.2$) cloud. 
This trend is a clear demonstration that the star formation in a more 
strongly magnetized cloud is retarded to a larger extent. The behavior 
of star formation in critical clouds is 
intermediate between those of subcritical and supercritical clouds 
discussed in the previous subsections. 

\section{Dense Cores in Standard Simulation}

Dense cores are the basic units of low-mass star formation. The 
best observed cores are those selected from nearby dark clouds 
of dispersed star formation that are large enough for detailed 
mapping. Individual cores in cluster-forming regions such as $\rho$ 
Oph tend to be smaller and thus less well resolved (Motte, Andre 
\& Neri 1998). In this section, we will concentrate on the cores formed in 
our standard simulation, which we believe is representative of the 
clouds undergoing dispersed star formation of low efficiency. 

In our standard simulation, dense cores are formed in supercritical 
filaments which, in turn, are produced in strongly shocked regions 
where ambipolar diffusion is accelerated. Since large-scale converging
motions are canceled out in post-shock regions, we expect the cores
to develop in a relatively quiescent environment (e.g., Padoan et al. 
2001). On the other hand,
the cores and the filaments in which they are embedded are pushed 
around constantly by the ambient turbulent flow and occasionally by 
protostellar outflows, so perfectly quiescent cores are not to be 
expected. Furthermore, inside individual cores, self-gravity can in 
principle generate substantial infall motions. The velocity fields 
of some starless cores have been mapped and 
more will become available in the ALMA era. They provide a contact 
between theory and observations. 

There is no unique definition of dense cores. Observationally derived 
properties of cores depend on the molecules or dust emission used 
to trace them. We adopt the following procedure to identify a core. 
First, we locate all local maxima of the column density distribution 
above a threshold $\Sigma_c = 10 \Sigma_0$, where $\Sigma_0$ is the
average column density. Around each maximum, we then draw a set of 
density contours and identify the region enclosed within a level 
$\Sigma_e=6\ \Sigma_0$ as a dense core. The boundary value $\Sigma_e$ 
is chosen to be roughly half of the peak densities of those cores 
that are barely above the threshold $\Sigma_c$. It corresponds to a 
visual extinction of magnitude 6 (for the canonical choice of average 
$A_V=1$), typical of many cores of dark clouds selected from the 
Palomar Sky Atlas (Myers et al. 1983). 

Cores are created, distorted and destroyed continuously. For 
illustration, we will concentrate on those identified at $t=2\ 
t_g$, the middle point of our standard simulation. By this time,
the initial turbulent velocity field has more or less relaxed.
We identified 10 cores in total (excluding a partial ring created 
by the outflow of a newly formed star). There are double-cores 
sharing a common envelope; they are split into two single cores. 
The cores are plotted and numbered in Fig.~9, and shown in greater
detail in Fig.~10. Plotted in each panel of Fig.~10 are the (column) 
density contours and vectors of the velocity field {\it relative 
to the density peak}. The relative velocity is a measure of the 
(nonthermal) velocity dispersion of the core. Some properties 
of the cores are summarized in Table~2. In this table, the core 
radius (column 3) is evaluated as the radius of a circle enclosing 
the same area as the $\Sigma_e=6\ \Sigma_0$ contour; for double
cores, the area is split evenly between the two components. The 
average flux-to-mass ratio in column 6 is defined as the total 
magnetic flux divided by the core mass, and the velocity dispersion 
in column 7 as the mass-weighted rms value of the flow velocity 
relative to the density peak. 

From Fig.~10 and Table~2, we find that the majority of cores have 
a subsonic internal velocity dispersion. One exception is core 
No.2 (panel~b in Fig.~10). 
Its peak density of $118\ \Sigma_0$ is the highest of all cores. 
The core has already collapsed, with a ``stellar'' mass of $0.13\ 
T_{10}^2/A_V\ M_\odot$ located within the highest density cell. 
In effect, it contains a ``Class 0'' protostar still actively 
accreting mass from the surrounding massive envelope (Andre et 
al. 2000); the large velocity dispersion of the core is thus 
gravitational in origin. Its mass accretion is terminated in another 
$\sim 7\times 10^4 T_{10}^{1/2}/A_V$~yrs, comparable to the duration
of the Class 0 phase. The core that most closely resembles the 
collapsed core is core No.6 (panel~g). It has the 
second highest peak density 
of $41.2\ \Sigma_0$. Systematic infall has already developed 
in this core, although still at a subsonic speed. It is well on 
its way to star formation, formally producing a ``finished'' star 
in about $0.18\ t_g=0.34\ T_{10}^{1/2}/A_V$~Myrs. The infall motions 
are in contrast to the systematic outward motions in core No.1 
(panel~a). 
This core has a peak density ($11.3\ \Sigma_0$) slightly above 
the threshold $\Sigma_c=10\ \Sigma_0$. It is in the process of 
dispersing away. There are three other dispersing cores. They 
are cores No.4, 5 and 10 (panels d, e and l, respectively). 
These cores were recently formed out of 
material swept up by protostellar outflows, and tend to have 
relatively high internal velocity dispersions. The dispersing 
cores also have higher-than-average flux-to-mass ratios, with 
values between $0.81$ and $1.03$. The high level of turbulence 
and high degree of magnetization are probably the reasons why 
these cores are dispersing rather than condensing.  

The core with the lowest degree of magnetization and the lowest 
velocity dispersion is core No.9 (panel~k). This roundish core 
has the smallest mass and radius of all cores, and is apparently 
close to a mechanical equilibrium. It lasts for nearly $0.45\ t_g$ 
(or about 5 times the local collapse time), before being induced 
to collapse by an protostellar outflow. 
 
The cores No. 7 (panel~i) and 8 (panel~j) are enclosed within a common 
envelope. Both have relatively high flux-to-mass ratios (0.91 and 0.80, 
respectively), which is probably the reason why they lived for a 
relatively long time (compared to their local 
collapse times) before merging and forming stars around $t=2.36\ 
t_g$. The longest living core is core No.3. It can be traced to 
the end of simulation, for more than 20 times its local collapse 
time. At the time $t=2\ t_g$ (shown in panel~c), the core has a 
relatively large velocity dispersion (close to the sound speed) and is 
being compressed by a converging flow. The compression was not able 
to push the core directly into collapse, probably because of its 
relatively large flux-to-mass ratio (0.88). This elongated core 
appears to be initially far out of equilibrium. It gradually 
relaxes toward a round configuration. At later times, a rapid 
rotation develops. Rotational support is probably the main reason 
for its extraordinary longevity. 

Panels (f)-(h) of Fig.~10 illustrate the evolution of the collapsing 
core No.6. As is typical, this core is initially elongated, as can
be seen in panel~(f) (at $t=1.78\ t_g$), when it is assembled by a 
converging flow. 
At this early stage of formation, the core (as outlined by the 
density contour at $\Sigma_e=6\ \Sigma_0$) is extended and turbulent, 
with a supersonic velocity dispersion. It becomes rounder and more 
quiescent at later 
times, with a velocity dispersion reaching down to $30\%$ of the 
sound speed at $t=2\ t_g$ (shown in panel g). The slow, subsonic 
contraction observed at this time soon gives way to a more rapid, 
supersonic collapse, which is shown in panel (h) (at $t=2.06\ t_g$). 
The dip in 
velocity dispersion appears to be a common feature of star-forming 
cores during the period after the external compression is relaxed 
but before dynamical collapse sets in. 
From panel (f) to (g), the core radius decreases by about a factor
of two. The fact that this core is 
more compact at later times is consistent with the H$^{13}$CO 
observations of Mizuno et al. (1994), which show a tendency for more 
evolved cores to be smaller in size. Note also that the density peak 
is offset from the geometric center of the core at all three times. 
Such a cometary shape is often observed in dense cores (Crapsi et al. 
2005). 

In summary, of the 10 cores shown in Fig.~9, one formed a star already 
(No.2), another is well on its way to star formation (No.6). After 
being strongly perturbed, three more cores collapse to form stars (No.7, 
8, and 9). Of the remaining 5 cores, four expanded and disappeared as 
their peak densities drop below the threshold for a core (No.1, 4, 5, 
and 9). The fate of the last core (No.3) is uncertain. 

\section{Turbulence Dissipation}
\label{decay}

Supersonic turbulence is known to dissipate quickly. This is also true in 
our simulations. The turbulence decay is illustrated in Fig.~11, where 
we plot the mass-weighted rms Mach number of the gas  
\begin{equation}
{\cal M}=\left[{\int\int \Sigma\ (V_x^2+V_y^2)\ dx\ dy\over \int\int 
\Sigma\ C_s^2\ dx\ dy} \right]^{1/2}
\end{equation}
of the standard simulation as a function of time. The Mach number drops 
from the initial value 10 to $\sim 4$ in one turbulence crossing time 
($t_{\rm x}=0.5\ t_g$). The decline slows down at later times, with a factor 
of two reduction from one to eight crossing times. The slowdown results 
from a decrease in shock strength, as the flow speed approaches the 
fast magnetosonic speed of the cloud. As a result of magnetic cushion, 
moderately supersonic motions can survive for a long time, especially
in relatively low density regions. 

For comparison, we also plotted the rms Mach number of the stars in 
Fig.~11. Except for an initial dip, the rms stellar Mach number 
increases gradually from $\sim 2$ to $\sim 3.5$, before dropping 
back to $\sim 3$. The Mach number is somewhat higher than that of 
the gas at late times. The most likely reason for this difference,
we believe, is that stars experience the full gravity of all the 
gas inside the simulation box, whereas the gas feels only a 
magnetically diluted gravity. 

Protostellar outflows appear to have little effect on the replenishment 
of turbulence in our standard simulation. When the outflow strength is 
reduced by a factor of 10, the rms Mach number of the gas is hardly 
changed (see solid lines in Fig.~12). The ineffectiveness stems from 
the slow rate of 
star formation in this moderately subcritical cloud, where a relatively 
small amount of momentum is injected into the cloud over a long period 
of time. Even in the case of marginally supercritical cloud, where the 
star formation rate is much higher, the dissipated turbulence is not 
replenished significantly, as shown in Fig.~12 (dotted lines). Only in 
the non-magnetic ($\Gamma_0=0$) and strong outflow ($f=0.1$) case, the 
rate of star formation is high enough that momentum injection from 
outflows can roughly offset dissipation and maintain the turbulence at 
a level comparable to the initial value for a collapse time. The high 
level of turbulence is also 
evident in the velocity field plotted in the first three panels of 
Fig.~7. When the outflow 
parameter $f$ is reduced to $0.01$, the turbulence in the non-magnetic
cloud decays with little replenishment, despite an increase in the rate 
of star formation. 

\section{Discussion}

\subsection{Dispersed Star Formation of Low Efficiency}

Magnetic regulation was originally envisioned for the dispersed mode 
of inefficient star formation (Shu et al. 1987). The best studied
example of this mode is the Taurus molecular cloud complex. With a 
mass $\sim 2\times 10^4$~M$_\odot$ (Ungerechts \& Thaddeus 1987) 
and $\sim 200$ low-mass young stellar objects (Kenyon \& Hartmann 
1995), the overall star 
formation efficiency is about $1\%$. Palla \& Stahler (2002) 
analyzed the star formation in Taurus both in space and in time. 
They found that it started $\sim 10$~Myrs ago at a low level, and 
the pace of star formation accelerated in the last few million 
years. Since the clouds must be at least as old as the oldest stars, 
Palla \& Stahler concluded that the Taurus clouds are probably 
older than $\sim 10$~Myrs. If true, the clouds would have lived 
for at least 5 times the gravitational collapse time $t_g$, since 
their average visual extinction $A_V\sim 1$ (Arce \& Goodman 1999) 
and temperature $T\sim 10$~K yield $t_g\sim 2$~Myrs. 

Hartmann (2003) questioned the above estimate of Taurus cloud lifetime. 
He pointed out that the oldest stars in Palla \& Stahler sample tend 
to be stars more massive than $\sim 1 M_\odot$ and that the ages of 
such stars are affected by uncertainties in birthline location. If 
these stars are discarded, the ages of the oldest stars (and thus 
the lower limit to the cloud lifetime) would be closer to 5~Myrs 
(or $\sim 2.5 t_g$) than to 10~Myrs. The shorter lifetime was taken 
as evidence against the traditional, quasi-static picture of magnetically 
regulated star formation. In the presence of a supersonic turbulence,
significant star formation can occur in moderately subcritical clouds 
in a few collapse times, so that even the shorter cloud lifetime is 
not a problem for the picture of star formation accelerated by 
turbulence and regulated by magnetic fields through ambipolar diffusion 
(for example, the SFE of the standard model reaches about 1 \% in 1-1.5 
$t_g$ or 2-3~Myrs for $\Gamma_0 = 1.2$ and ${\cal M}=10$). 

Hartmann (2003) suggested that star formation in Taurus is rapid and 
dynamic. A specific scenario envisioned was molecular cloud formation 
and prompt star formation in converging flows of atomic clouds 
(Hartmann et al. 2001). It remains to be demonstrated, however, that 
only one percent of the molecular gas can be turned into stars in this
scenario. Numerical simulations of large-scale turbulent flows 
containing many Jeans masses invariably show rapid conversion of gas 
into stars, with efficiencies much higher than one percent (e.g., Bate 
et al. 2003; see also section \S~\ref{supnon}). The high efficiency 
problem is one of the main factors that motivated the original 
proposition that star formation in regions like Taurus is magnetically 
regulated (Shu et al. 1987).

The combination of a strong magnetic field and supersonic turbulence 
allows for a range of possibilities. In our standard simulation, the 
cloud is moderately subcritical (with a dimensionless flux-to-mass
ratio $\Gamma=1.2$) and the turbulence is relatively strong (with a 
rms Mach number ${\cal M}=10$). By the end of 4 gravitational 
collapse times ($t=4\ t_g=7.60\ T_{10}^{1/2}/A_V$~Myr), 5.3\% of the 
cloud mass has been converted into stars. It corresponds to an average 
rate of star formation per unit mass   
$
7\times 10^{-9} {A_V/T_{10}^{1/2}}
$
yr$^{-1}$. Interestingly, this (specific) rate is not far from the 
Galactic average of $\sim 
3 \times 10^{-9}$~yr$^{-1}$ for a star formation rate of $3 M_\odot$ 
per year inside the solar cycle (Scalo 1986) and a molecular mass of 
$10^9 M_\odot$ in that region (Blitz \& Williams 1999). One can make 
the star formation more inefficient by increasing the magnetic field 
strength and/or reducing the turbulent Mach number. Reducing the Mach 
number of the standard simulation from 10 to 3, for example, lowers 
the SFE at $t=4\ t_g$ from 5.3\% to 2.0\%. The reduction factor is 
higher at earlier times, when few stars are formed in the lower Mach 
number case. The pace of its star formation accelerates at later times 
(see Fig.~8), as inferred by Palla \& Stahler (2002). 

To summarize, we believe that dispersed star formation of low efficiency
in regions like the Taurus molecular clouds naturally occurs in 
moderately magnetically subcritical clouds, through ambipolar diffusion 
that is accelerated by turbulence. Whether the bulk of the cloud material 
is indeed magnetically supported remains to be determined through Zeeman 
measurements. Crutcher \& Troland (2000) measured the line-of-sight 
field strength toward the L1544 region in Taurus. The inferred 
deprojected flux-to-mass ratio $\Gamma \sim 0.5$ is in agreement with 
core formation models involving ambipolar diffusion (Basu \& Mouschovias 
1994; Nakamura \& Li 2003; see the ratios for collapsing cores in 
column~6 of Table~2). For the 
general background of the Taurus clouds where $A_V \sim 1$ (Arce \& Goodman 
1999), the critical field strength is $\sim 7.6$~$\mu$G. The line-of-sight 
component is expected to be smaller than this value by a typical factor 
of $\sim 2-4$ (depending on the cloud geometry; Shu et al. 1999). 
Whether such a weak field can be directly measured remains to be seen. 
Wolleben \& Reich (2004) inferred a field strength exceeding $\sim 20 
\mu$G from modeling 
polarization measurements at 21 and 18 cm toward the Taurus cloud 
complex. The value need confirmation from Zeeman Measurements. 

\subsection{Quiescent Cores and Inefficient Star Formation}

Starless cores define the initial conditions for low-mass star formation 
(Andre et al. 2000). Their structure provides clues to how they are 
formed. Dust continuum observations have shown that these cores have 
flat-topped density profiles (Ward-Thompson et al. 1994). This feature 
can, however, be explained by essentially any type of core formation 
scenario. More discriminating are molecular line observations, which 
are sensitive to the internal motions of cores. Lee et al. (1999) and 
Gregersen \& Evans (2000) carried out line surveys of nearby starless 
cores. A general conclusion is that optically thick lines can be displayed 
either to the blue or red side of optically thin lines, with an 
overabundance of sources with blue asymmetry. The simplest interpretation 
is that cores with ``blue'' asymmetry are contracting, and those with 
``red'' asymmetry are expanding. 
Both collapsing and dispersing cores are observed in our simulations. 
Our collapsing cores tend to have higher column densities than the 
dispersing cores (see Table~2 for examples). This trend is consistent 
with the finding of Lee et al. (1999;  see also 
Gregersen \& Evans 2000) that the best infall candidates tend to 
have the highest column densities. 

There are two important characteristics for the dense cores formed in 
our standard simulation, which is representative of inefficient, 
dispersed star formation in moderately magnetically subcritical 
clouds. First, most cores are supercritical. This is the case, for 
example, for 9 out of the 10 cores shown in Fig.~\ref{fig:9}; the 
remaining one is very close to critical, with an average flux-to-mass 
ratio $\Gamma=1.03$. The dearth of
subcritical cores is in agreement with Nakano (1998), who argued 
that such cores are unlikely to exist because they  
require a strong external confinement to stay in equilibrium. Even 
if external confinement is available, it would be difficult to 
maintain the observed level of nonthermal velocity dispersion 
for a significant
fraction of the core's (long) lifetime. These difficulties have led 
Nakano (1998) to abandon the picture of core formation through 
ambipolar diffusion in favor of one involving turbulence decay. We 
believe the abandonment is premature. We have 
shown that converging flows in a strong turbulence can create and 
confine high density subcritical regions long enough for accelerated 
ambipolar diffusion to produce supercritical material, particularly 
in {\it moderately subcritical clouds} where the magnetic forces that 
drive the expansion are already offset by self-gravity to a large 
extent (so that the confinement requirement is much reduced; Shu 
\& Li 1997) and a relatively small reduction in magnetic flux can 
make a piece of cloud supercritical. Once created, supercritical cores 
can be held together by self-gravity and are able to collapse to form 
stars more quickly than externally-confined subcritical cores (so
that excessive turbulence decay is less of a problem). In our picture, 
the bulk of magnetic flux reduction occurs dynamically (in shocks), 
rather than statically, through ambipolar diffusion when the cloud 
is strongly turbulent. As the level of turbulence decreases, 
quasi-static ambipolar diffusion becomes more significant. 

The second important characteristic of the cores is that they have 
predominantly subsonic internal motions (see Table~2 for examples). 
This is in agreement with the finding that the velocity dispersions 
of starless cores in regions of inefficient, dispersed star formation 
such as the Taurus clouds tend to be thermally dominated (Jijina et 
al. 1999; Lee et al. 1999). Quiescent cores are difficult to arrange 
in turbulent, non- or weakly-magnetized clouds that contain many Jeans 
masses: these cores expand or collapse dynamically unless the core 
mass happens 
to be close to the local thermal Jeans mass. In our simulations, 
quiescent cores appear to form out of the turbulent, moderately 
subcritical background without fine tuning. The reason, we believe, 
is that our cores, 
although supercritical in general, remain strongly magnetized, with a 
dimensionless flux-to-mass ratio $\Gamma$ close to unity (the critical 
value; see Table~2). Roughly speaking, the magnetic tension force is 
$-\Gamma^2$ times the self gravity (Shu \& Li 1997, their equation 
[2.16]; the ratio would be exact if $\Gamma$ is spatially constant; 
see also Nakano 1998). Inside each core, the self gravity is balanced
out by the magnetic tension force to a large extent (the ``effective'' 
gravity is $1-\Gamma^2$ times the undiluted value). Barring occasional 
strong external perturbations, the magnetically levitated cores tend 
to contract or expand in ``slow motion'' regardless of the amount of 
mass involved. The slower motions enable the quiescent cores to live 
longer than their dynamically evolving counterparts, which tend to 
increase their number at any given time. 
The long-lived cores that eventually form stars have masses greater 
than the local thermal Jeans mass but are gravitationally stable due 
to significant magnetic support. They tend to oscillate with subsonic 
speeds before rapid contraction. The oscillation, we speculate, may 
excite the type of pulsating flow pattern observed in B68 (Lada et al. 
2003). Such cores start to collapse either after merging with other 
supercritical regions or after significant additional flux loss due 
to ambipolar diffusion. As the level of turbulence decreases, the 
latter becomes increasingly more important. 

Magnetic levitation also lies at the heart of the low efficiency of star 
formation.  In a cloud that is subcritical overall, the self gravity is
completely offset by magnetic forces except in localized pockets---the 
supercritical ($\Gamma < 1$) filaments (see Figs.~2 and 4), which contain 
a fraction of the cloud mass (about 1/3 in our standard simulation). 
Only a small fraction of the supercritical material resides in dense 
cores (about 1/10 in our standard simulation) that are intimately 
related to star formation. Not all cores form stars; some of them 
disperse away without star formation. For those cores that do form 
stars, their efficiencies are further limited by protostellar outflows, 
which can unbind the majority of core material. 

Our picture of turbulence-accelerated, magnetically regulated star 
formation has implications on astrochemistry. Cores formed directly
out of the low-density background from rapid turbulent compression 
may have had little time for chemical reactions. They may be 
identified with chemically ``young'' cores such as L1521E that are 
rich in carbon-chain species like CCS and show little evidence of 
CO depletion (Hirota et al. 2002; Tafalla \& Santiago 2004). Some 
of the dense cores expand back to low densities, and it would be 
interesting to explore the effects of these dispersed cores (where
chemical depletion, for example, may have occurred) on the chemistry 
of the more diffuse region (Garrod et al. 2005). Others evolve 
further towards star formation, 
through continued ambipolar diffusion and/or external perturbations 
(such as merging and shock compression). They may be identified 
with more evolved cores such as L1544 and L1521F (e.g., Crapsi et 
al. 2004) that are thought to be on the brink of star formation. 
The interplay between turbulence and magnetic fields in the presence 
of ambipolar diffusion allows for a variety of possibilities for 
the evolutionary history of starless cores, which may be constrained 
through astrochemical modeling. 

\subsection{Implications on Cluster Formation}

The majority of stars are thought to form in clusters. One of the
best-studied nearby cluster-forming regions is the $\rho$ Oph 
core. It has an estimated stellar mass $53\ M_\odot$ and core mass  
$550\ M_\odot$, yielding a SFE of $\sim 9\%$ (Lada \& Lada 2003). 
This efficiency may increase with time, to a value of order $\sim 
30\%$, the maximum value inferred for nearby embedded clusters 
(Lada \& Lada 2003). SFEs of only a few tens of percent are not
easy to obtain in turbulent cores that are not magnetized, unless
the cores are constantly driven on a small enough scale (Klessen 
et al. 2000). In our simulations, the non-magnetic cloud collapses 
promptly, forming stars with an efficiency between $\sim 55\%$ to 
$70\%$ at 1.5 collapse times, despite the presence of protostellar 
outflows. The efficiency is reduced substantially as the field 
strength increases toward the critical value. The marginally 
supercritical cloud in our models has a field strength $20\%$ below 
the critical value.  Its star formation efficiency approaches 
$\sim 30\%$, close to the observationally inferred maximum value. 
When and how the star formation is actually stopped in a Taurus-like
or $\rho$ Oph-like cloud remains uncertain (e.g., Palla \& Stahler 
2002). 

The non-magnetic cloud that we modeled has another difficulty. It 
produces a mass distribution that is dominated by a network of dense, 
thin filaments, with large voids in 
between, similar to the structure obtained in pressureless cosmological 
simulations (see Fig.~7). Nearby cluster-forming regions such as 
$\rho$ Oph (Motte et al. 1998) and Serpens (Olmi \& Testi 2002) 
cores appear to have a smoother mass distribution, with less cloud 
mass concentrating in dense clumps than in inter-clump regions 
(Johnstone et al. 2004). This difficulty disappears in marginally 
supercritical clouds, where the dense cores are surrounded by 
extended, moderately dense envelopes. The better agreements in SFE 
and cloud mass distribution in the presence of a dynamically-important 
magnetic field lead us to believe that magnetic 
fields play a role even in cluster formation in the present-day 
Galaxy. This contention is consistent with available Zeeman 
measurements, which show that the (deprojected) flux-to-mass ratios 
are not far from the critical value in highly turbulent regions of 
high-mass star formation (Crutcher 1999). 

\subsection{Caveats and Future Work} 

Our calculations are two dimensional. The adopted sheet-like geometry 
is valid for magnetically-supported, multi-Jeans mass clouds that 
are subcritical or marginally supercritical but not necessarily for 
weakly magnetized clouds, unless they are flatten externally (by shocks, 
for example). Vertical force balance is assumed at all times, which 
may break down in shocked regions immediately following strong
compression. Such regions can puff up temporally, lowering the volume 
density (and thus the rate of ambipolar diffusion). The reduction may 
lead to a somewhat lower rate of star formation. Addressing this issue 
would require 3D calculations. 

3D calculations will also be needed to address the issue of magnetic
braking. Some of the cores formed in our simulations rotate rapidly. 
A particular example is the core No.3 shown in Fig.~9. It appears
to be rotationally supported at late times, lasting for more than 20 
times the local gravitational collapse time. The rapid rotation is 
expected to be reduced by magnetic braking (Basu \& Mouschovias 1994), 
which is not included in our simulation. The braking should reduce 
the rotational component of the velocity field inside a core in 
general, perhaps making the core more quiescent. The reduction factor 
will probably be less than that obtained by Basu \& Mouschovias for 
quasi-statically evolving clouds, since in our picture the core 
formation time is shortened by external compression. 

In our simulations, stars are taken to have the same final mass, which is
obviously an oversimplification. An alternative is to assume that
the time for a star to accrete its mass is fixed. Some justification 
for this assumption comes from the finding that the range in 
accretion time is much narrower than that in stellar mass (by an 
order of magnitude; Myers \& Fuller 1993). Resolving the accretion 
flow around each forming star would require a spatial resolution 
higher than that of our simulations. Future higher resolution 
simulations may allow us to directly tackle the problem of initial 
mass function. We have assumed for simplicity that the magnetic 
flux-to-mass ratio is initially uniform throughout the cloud. This 
assumption will be relaxed in future studies. 

The replenishment of turbulence is a potential problem. In our standard
simulation of inefficient star formation in a magnetically subcritical 
cloud, the velocity dispersion drops quickly in the first turbulence
crossing time. The decline rate slows down at later times, with the 
flow velocity approaching twice the sound speed for the cloud as a whole; 
it is somewhat higher in the lower density regions, where large patches 
of more or less coherent velocity field exist. More organized velocity 
fields tend to be less dissipative, as suggested by Palla \& Stahler 
(2002). The typical flow velocity of a few times the sound speed is not 
as large as the velocity dispersions observed on the scale of cloud 
complexes. How the large-scale turbulence is generated and maintained 
remains uncertain (see Elmegreen \& Scalo 2004 for a review).  

\section{Summary}

We have explored numerically the interplay of magnetic fields and 
turbulence in the presence of ambipolar diffusion in a sheet-like
geometry, taking into account protostellar outflows in an 
approximate manner. In strongly turbulent, moderately magnetically 
subcritical clouds, we found a dispersed mode of star formation 
with efficiencies at several percent level. Dense cores formed in 
such clouds tend to be quiescent, with subsonic internal motions, 
in agreement with observations. For strongly turbulent, marginally 
supercritical clouds, the efficiency increases to a few tens of 
percent, which is in the range inferred for nearby cluster-forming 
regions. If not regulated by magnetic fields at all, the efficiency 
of star formation exceeds half in a collapse time, despite vigorous
protostellar outflows. We conclude that a happy marriage between
turbulence and magnetic fields must be sought, particularly for 
the inefficient, dispersed mode of star formation, perhaps for 
the more efficient, clustered mode as well. Further refinements 
of this picture of turbulence-accelerated, magnetically regulated 
star formation require 3D simulations. 

\acknowledgments 
Support for this work was provided by Grant-in-Aid for Scientific 
Research (No. 15740117) by the Ministry of Education, Culture, 
Science and Technology of Japan and NSF grant AST-0307368.

\clearpage

\begin{deluxetable}{lllllll}
\tablecolumns{7}
\tablecaption{Model Parameters}
\label{tab:model}
\tablehead{
\colhead{Model} & \colhead{$\Gamma_0$} & \colhead{$\cal{M}$} & 
\colhead{$n$} & \colhead{$f$} & Turbulence $^a$ & 
\colhead{Notes}  }
\startdata
S1 & 1.2 & 10 & 3 & 0.1 & A & standard model, subcritical\\
S2 & 1.2 & 10 & 3 & 0.01 & A & weak outflow \\
R1 & 1.2 & 10 & 3 & 0.1 & B & different random realization \\
R2 & 1.2 & 10 & 3 & 0.1 & C & different random realization \\
I1 & 1.2 & 10 & 3 & 0.1 & D & different random realization \\
I2 & 1.2 & 10 & 3 & 0.1 & E & different random realization \\
N1 & 1.2 & 10 & 5 & 0.1 & F & different power spectrum \\
N2 & 1.2 & 10 & 1 & 0.1 & G & different power spectrum \\
U1 & 0.8 & 10 & 3 & 0.1 & A & marginally supercritical\\
U2 & 0.8 & 10 & 3 & 0.01 & A & supercritical, weak outflow  \\
O1 & 0.0 & 10 & 3 & 0.1 & A &  non-magnetized \\
O2 & 0.0 & 10 & 3 & 0.01 & A & non-magnetized, weak outflow \\
S3 & 1.2 & 3 & 3 & 0.1 & A & intermediate turbulence \\
S4 & 1.2 & 1 & 3 & 0.1 & A & weak turbulence \\
C1 & 1.0 & 10 & 3 & 0.1 & A & magnetically critical \\
C2 & 1.0 & 3 & 3 & 0.1 & A & critical, intermediate turbulence \\
C3 & 1.0 & 1 & 3 & 0.1 & A & critical, weak turbulence\\
\enddata
\tablenotetext{a}{The letter A denotes the random realization of 
turbulent velocity field used in the standard simulation, and B, 
C, D and E for different random realizations of the same power 
spectrum as A. The velocity fields in D and E are incompressible. 
The power spectrum in E (F) is steeper (shallower) than that in
A-E. }
\end{deluxetable}

\begin{deluxetable}{llllllll}
\tablecolumns{8}
\tablecaption{Properties of the Cores Identified in the Standard 
Simulation at $t=2\ t_g$ $^a$}
\label{tab:core}
\tablehead{
\colhead{Core} & \colhead{Core}& \colhead{Core} & \colhead{Peak} & 
\colhead{$\Gamma$ at} & \colhead{Average} & 
\colhead{Velocity}  & \colhead{ Notes on Core} \\
\colhead{No.} & \colhead{Mass} & \colhead{Radius}& \colhead{$\Sigma$} & 
\colhead{$\Sigma$ Peak} & \colhead{$\Gamma$} & 
\colhead{Dispersion}  
& \colhead{Evolution}  }
\startdata
1 & 0.32 & 0.11 & 11.3 & 0.85 & 0.89 & 0.48 & expanding, dispersed\\
2 & 0.87 & 0.15 & 118 & 0.60 & 0.67 & 1.40 & collapsing, star formed at t=2.04 $t_g$ \\
3 & 0.32 & 0.10 & 18.0 &0.79 & 0.88 & 0.97 & fast rotation develops, fate unknown\\
4 & 0.48 &0.12 & 21.2 &0.82 & 0.85 & 1.41 & expanding, dispersed\\
5 & 0.35 &0.12 & 12.9 &0.77 & 0.81 & 0.64 & expanding, dispersed\\
6 & 0.91 &0.17& 17.4 & 0.85 & 0.90 & 1.33 & core No. 6 at t=1.78 $t_g$   \\
6 & 0.34 &0.095&41.2 &0.57 & 0.61 & 0.30 & collapsing, star formed at t=2.18 $t_g$ \\
6 & 0.33 & 0.077&90.2 & 0.53 & 0.78 & 1.18 & core No. 6 at t=2.06 $t_g$   \\
7 & 0.32 & 0.13& 11.8 & 0.85 & 0.91 & 0.55 & collapsing$^b$, star formed t=2.36 $t_g$ \\
8 & 0.42 &0.13&16.9 & 0.68 & 0.80 & 0.40 & collapsing$^c$, star formed t=2.36 $t_g$ \\
9 & 0.11 &0.068&11.4 & 0.53 & 0.57 & 0.15 & collapsing$^d$, star formed t=2.44 $t_g$ \\
10 & 0.54 &0.15&11.7 & 0.85 & 1.03 & 0.80 & expanding, dispersed\\
\enddata
\tablenotetext{a}{the units for the dimensionless quantities listed can be 
found in \S~\ref{units}.}
\tablenotetext{b}{this core merges with core No.8 to form a star.}
\tablenotetext{c}{this core merges with core No.7 to form a star.}
\tablenotetext{d}{this core collapses to form a star after colliding 
with a dense region located between cores No. 7 and 9.}
\end{deluxetable}

\begin{figure}
\epsscale{1.0}
\plotone{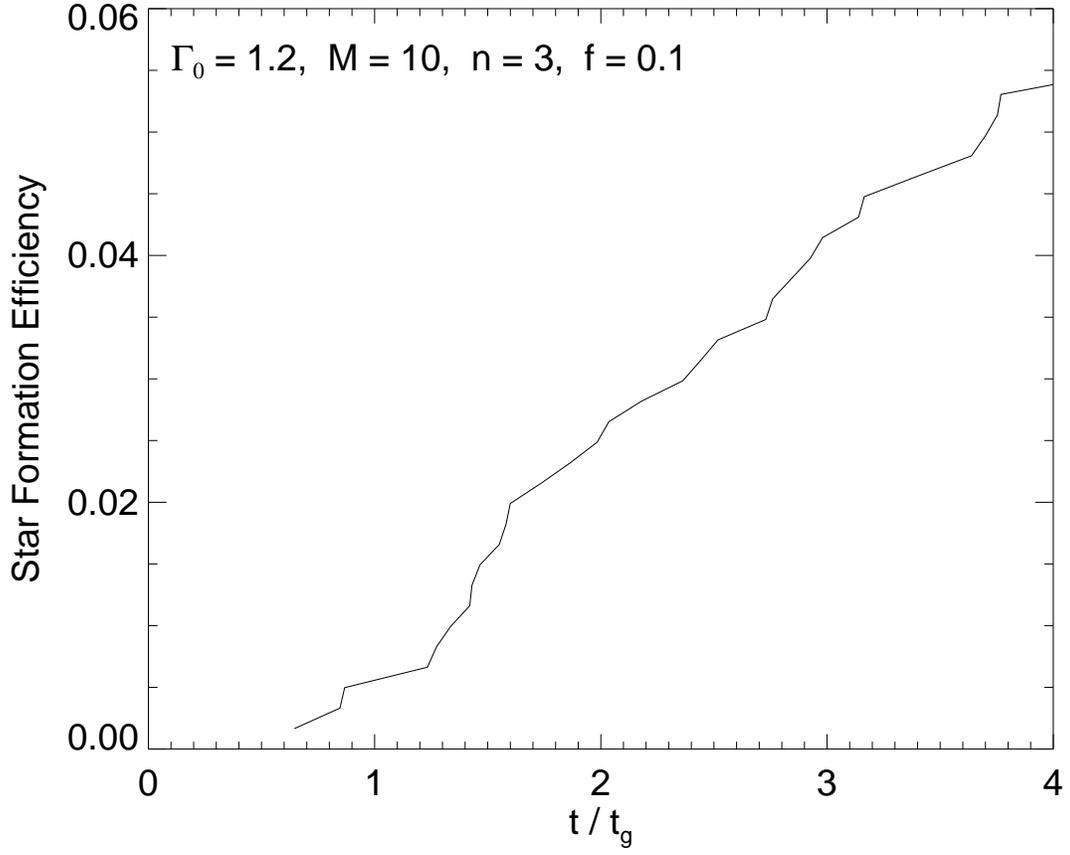}
\caption{The efficiency of star formation in the standard simulation 
(Model S1 in Table~1) as a function of time (in units of the 
gravitational collapse time $t_g=1.90\times 10^6 T_{10}^{1/2}/A_V$~yr).
 \label{fig:1}}
\end{figure}

\begin{figure}
\epsscale{1.0}
\plotone{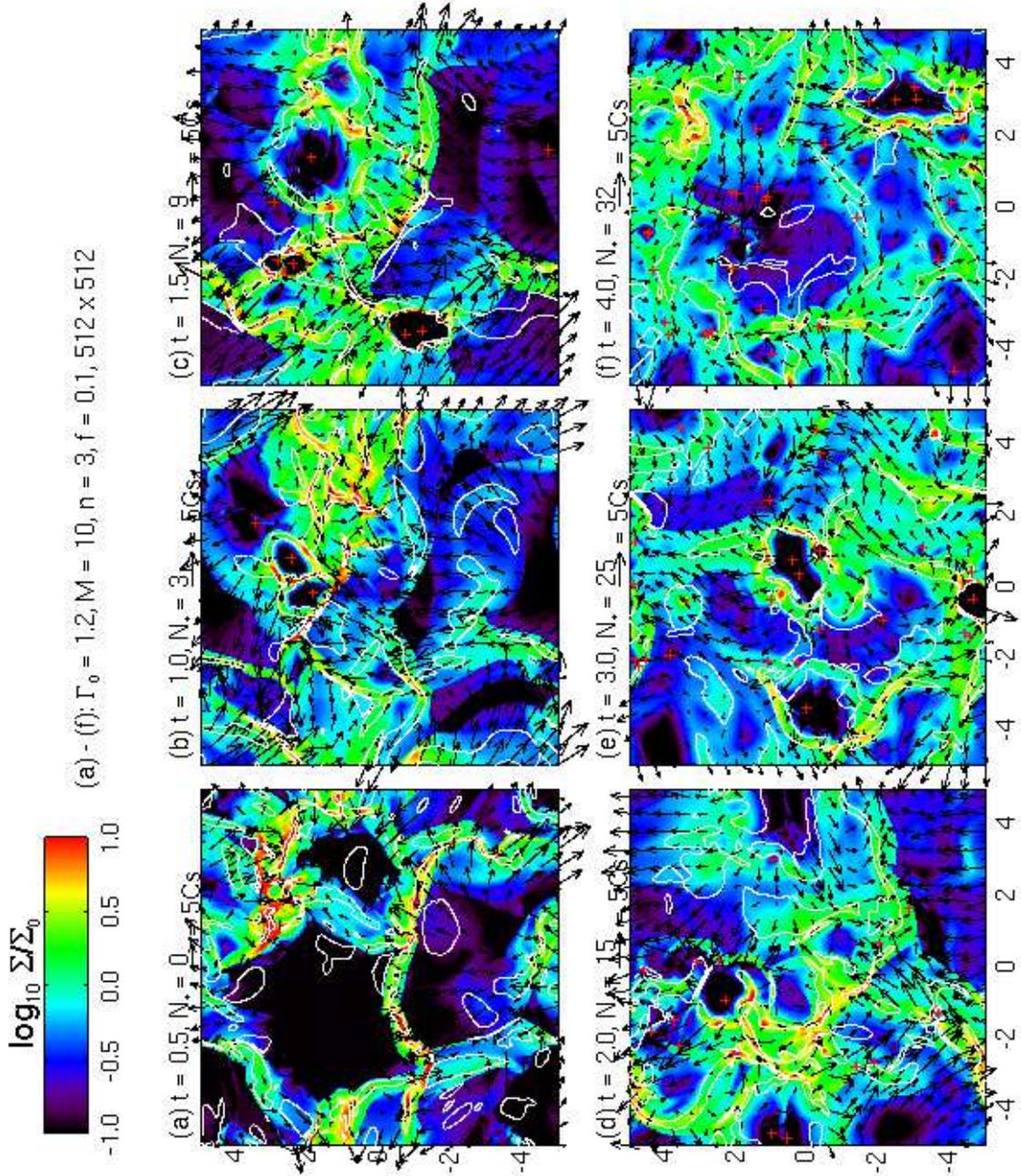}
\caption{Snapshots of the standard simulation at six representative 
times (given above each panel in units of $t_g$). Superposed on the 
color map of column density are velocity vectors (with normalization
shown above each panel, where $C_s$ is the isothermal sound speed), 
contours of critical flux-to-mass ratio, and locations of the stars 
formed (denoted by crosses). The number of stars $N_*$ is given 
above each panel. The simulation box has 10 Jeans lengths ($L_J=0.37\ 
T_{10}/A_V$~pc) on each side.  
\label{fig:2}}
\end{figure}

\begin{figure}
\epsscale{1.0}
\plotone{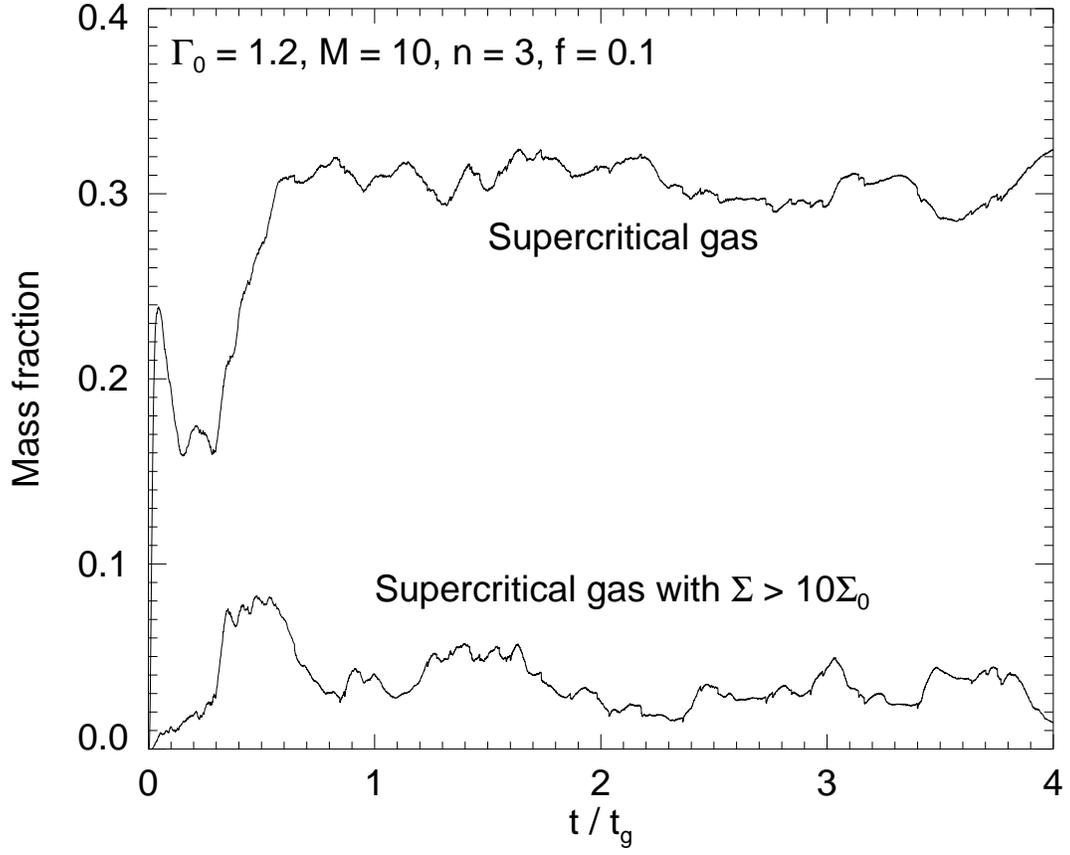}
\caption{Mass fraction of the magnetically supercritical material (upper
curve) created in the initially subcritical cloud of the standard simulation 
through ambipolar diffusion as a function of time. Only a fraction of 
this supercritical material resides in dense regions above ten times the 
average column density. The mass fraction of this dense, supercritical
material is shown as the lower curve. 
\label{fig:3}}
\end{figure}

\begin{figure}
\epsscale{1.0}
\plotone{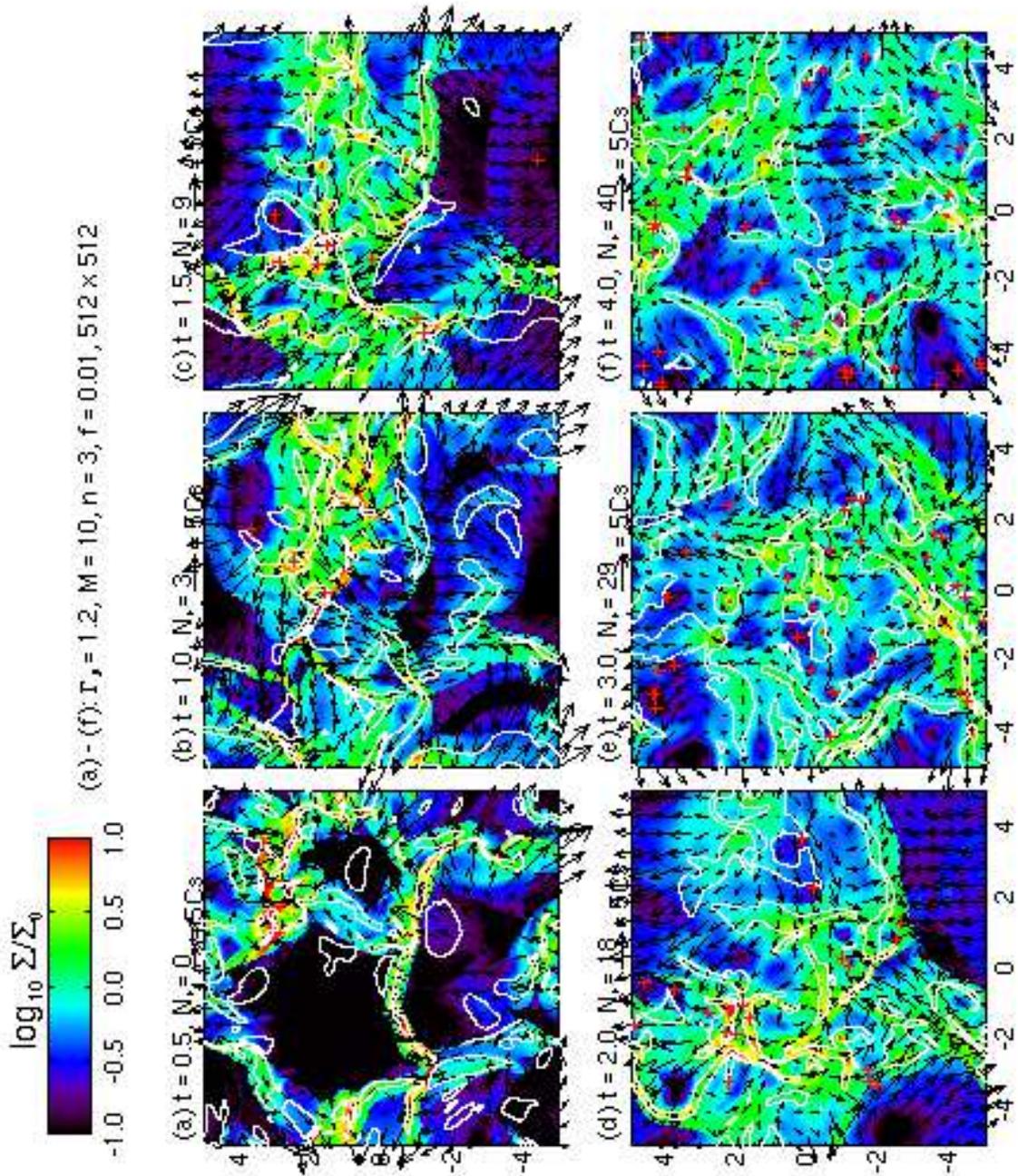}
\caption{Snapshots of a variant of the standard simulation with a weaker 
outflow of $f=0.01$ (Model S2 in Table~1). Superposed on 
the color map of column density are velocity vectors (with normalization
shown above each panel, where $C_s$ is the isothermal sound speed), 
contours of critical flux-to-mass ratio, and locations of the stars 
formed (denoted by crosses). The number of stars $N_*$ is given above 
each panel. The simulation box has 10 Jeans lengths 
($L_J=0.37\ T_{10}/A_V$~pc) on each side.  
\label{fig:4}}
\end{figure}

\begin{figure}
\epsscale{0.8}
\plotone{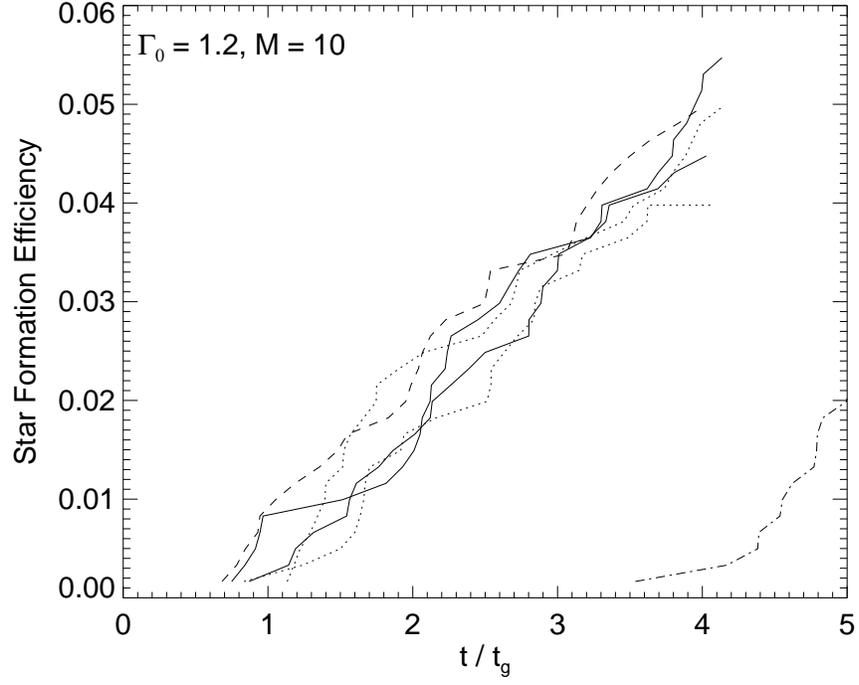}
\caption{Efficiencies of star formation of the six variants of the standard 
simulation listed in Table~1. The solid lines are for Models R1 and R2
that have the same power spectrum as the standard model, but different
random realizations. The dotted lines are for Models I1 and I2 whose
turbulent velocity fields are incompressible. The dashed (dash-dotted) 
line is for Model N1 (N2) that has a steeper (shallower) power spectrum 
than the standard model. 
 \label{fig:5}}
\end{figure}

\begin{figure}
\epsscale{1.0}
\plotone{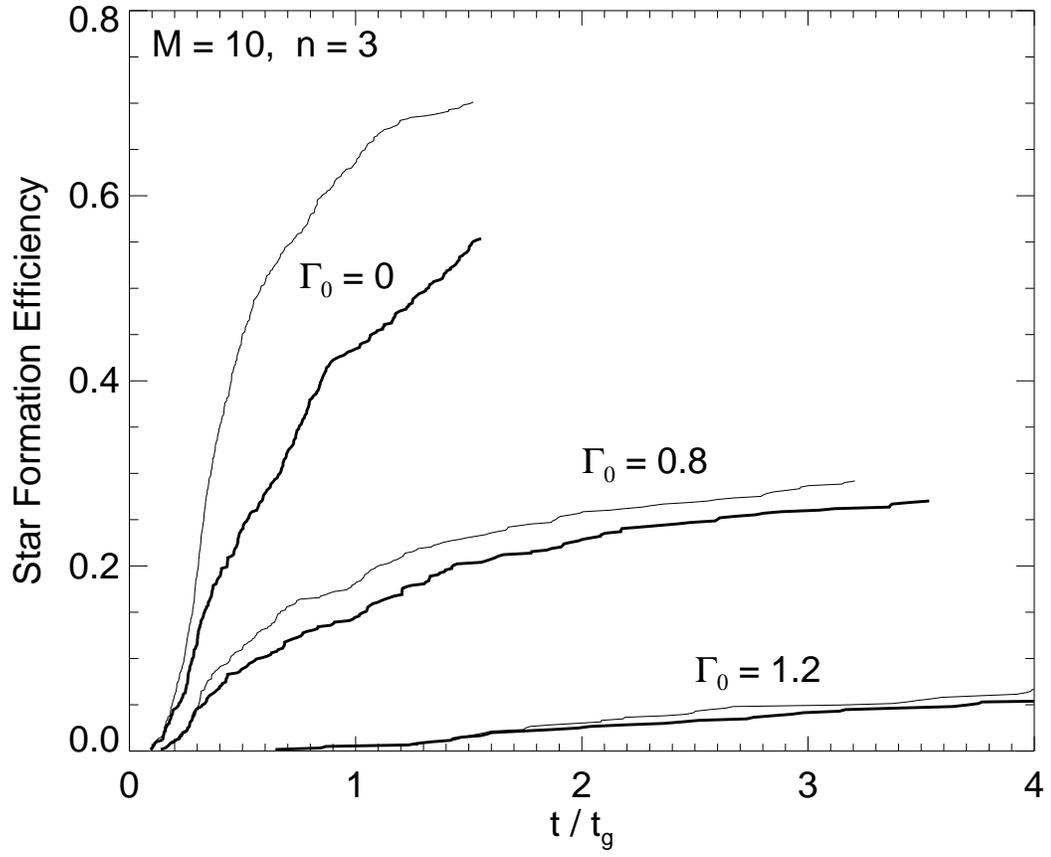}
\caption{Efficiencies of star formation for clouds of the same initial 
turbulent velocity field but different degrees of magnetization,
measured by the dimensionless flux-to-mass ratio $\Gamma_0$ labeled
beside the curves. The heavy (thin) solid lines are for strong (weak)
outflow cases with $f=0.1$ (0.01). 
\label{fig:6}}
\end{figure}

\begin{figure}
\epsscale{1.0}
\plotone{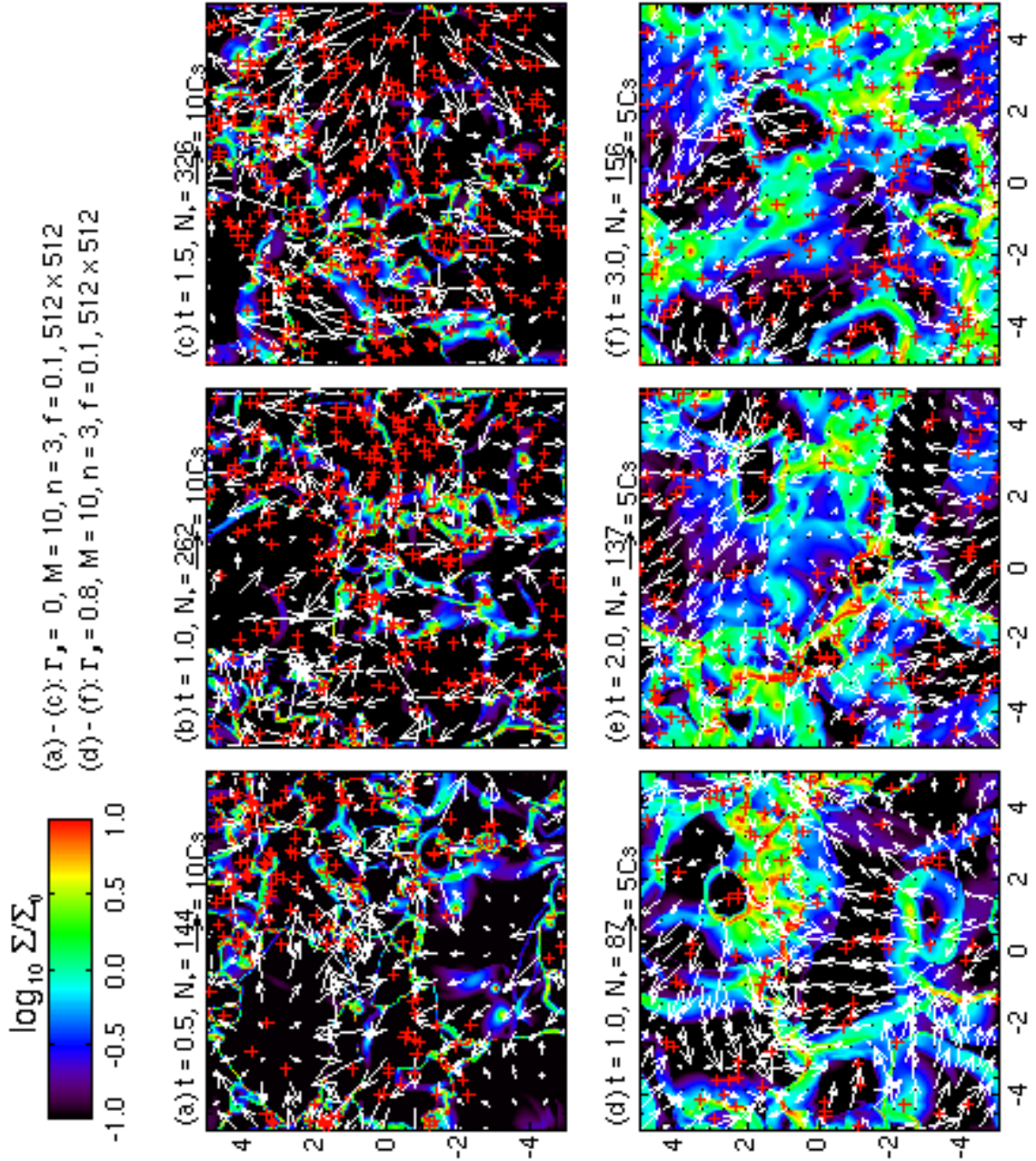}
\caption{Snapshots of a non-magnetic (Model~O1; panels a-c) and marginally 
supercritical cloud with $\Gamma_0=0.8$ (Model~U1; panels d-f) at three
representative 
times (given above each panel in units of $t_g$). Superposed on the 
color map of column density are velocity vectors (with normalization
shown above each panel, where $C_s$ is the isothermal sound speed)
and locations of the stars formed (denoted by crosses). The number 
of stars $N_*$ is given above each panel. The simulation box has 
10 Jeans lengths ($L_J=0.37\ T_{10}/A_V$~pc) on each side.  
\label{fig:7}}
\end{figure}

\begin{figure}
\epsscale{1.0}
\plotone{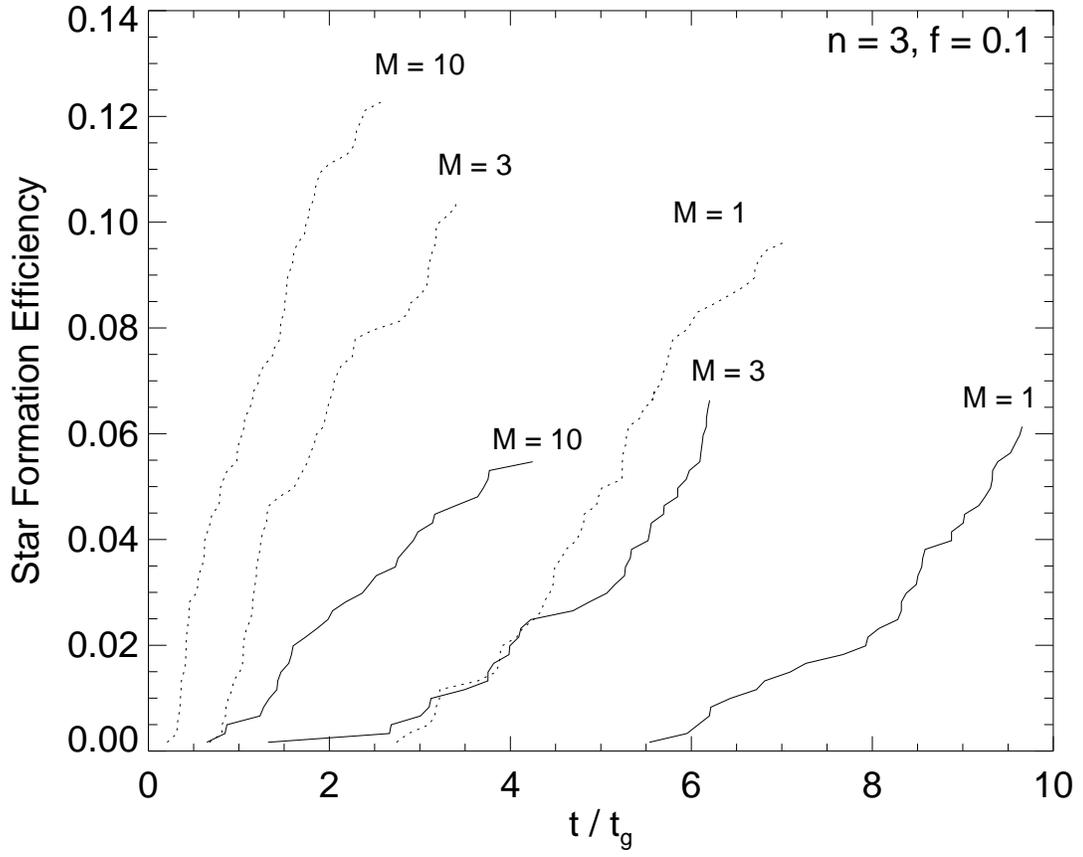}
\caption{Efficiencies of star formation for the standard subcritical 
cloud ($\Gamma_0=1.2$; solid lines) and a magnetically critical cloud 
($\Gamma_0=1$; dotted lines) for different values of turbulent Mach 
number ${\cal M}$. The role of strong turbulence in accelerating star 
formation is evident. 
\label{fig:8}}
\end{figure}

\begin{figure}
\epsscale{1.0}
\plotone{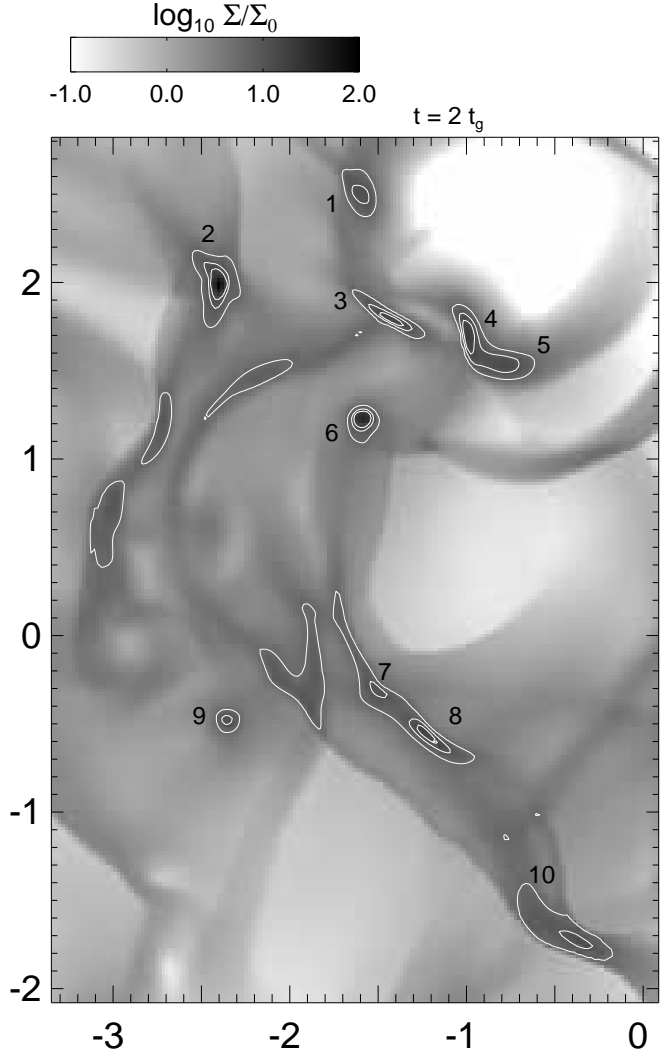}
\caption{Spatial distribution of the dense cores identified in the standard 
simulation at the time $t=2\ t_g$. The contours of column density are at
$\Sigma=6, 10$ and $14\ \Sigma_0$. Dense cores are enclosed by at least 
two contours. The core numbers are labeled. 
\label{fig:9}}
\end{figure}

\begin{figure}
\epsscale{0.95}
\plotone{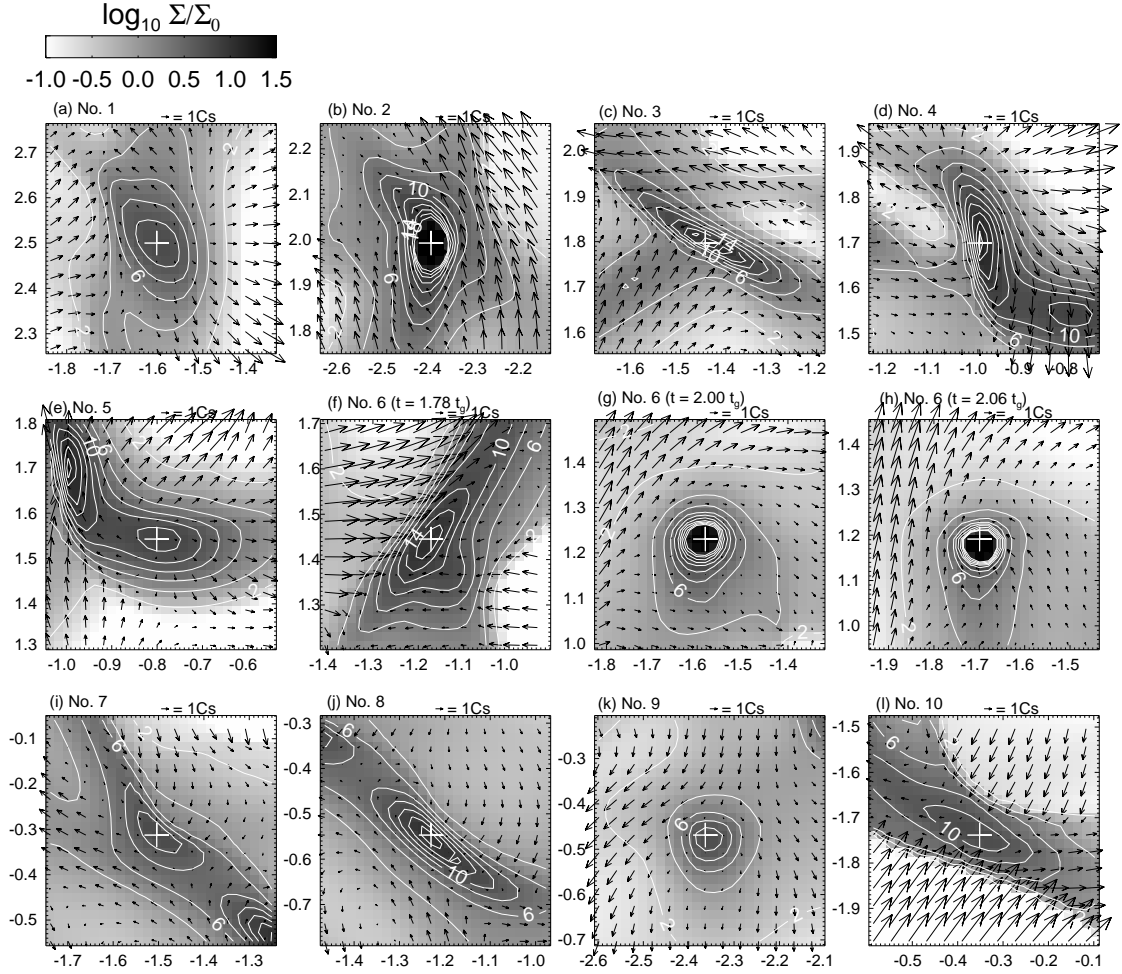}
\caption{Enlarged views of the dense cores shown in Fig.~\ref{fig:9},
except for panels (f) and (h), which display the core No.6 at two 
different times (labeled above the panels). Plotted in each panel are 
the contours of column density and vectors of the velocity relative
to the density peak (marked by a cross). The velocity normalization 
is shown above each panel, where $C_s$ is the isothermal sound speed. 
\label{fig:10}}
\end{figure}

\begin{figure}
\epsscale{0.8}
\plotone{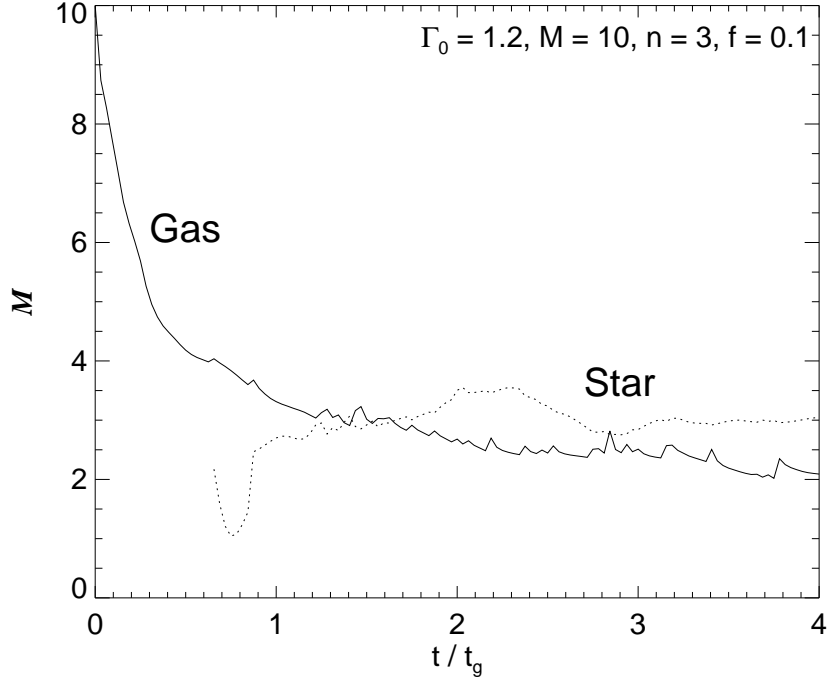}
\caption{The mass-weighted rms Mach numbers of the gas (solid line) and 
stars (dotted line) for the standard simulation as a function of time. 
Note the slowdown of turbulence decay at late times.  
\label{fig:11}}
\end{figure}

\begin{figure}
\epsscale{0.8}
\plotone{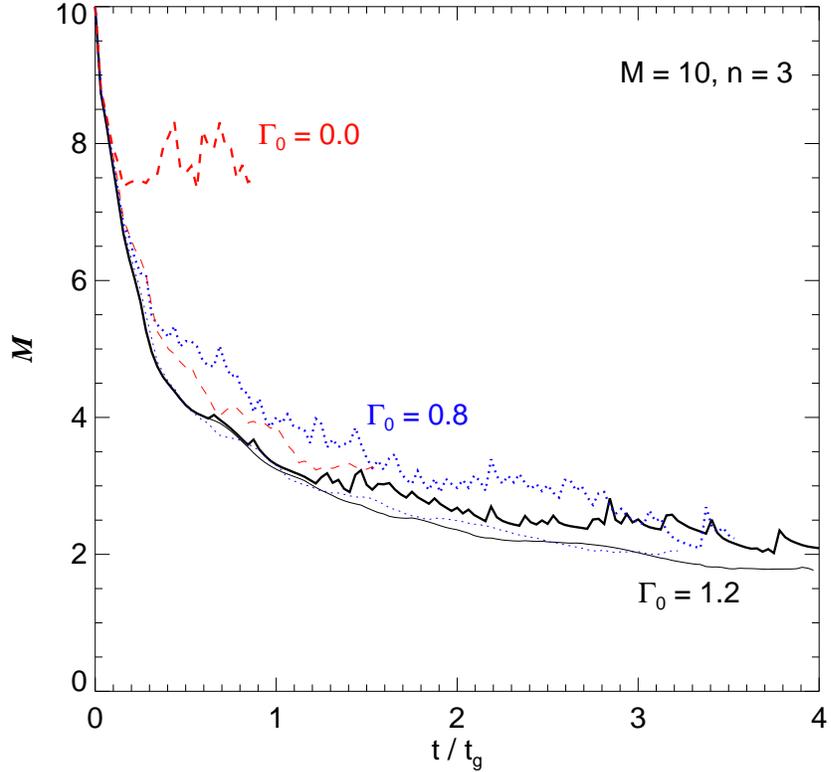}
\caption{The mass-weighted rms Mach numbers for the gas in the $\Gamma_0
=1.2$ (subcritical; solid lines), 0.8 (supercritical; dotted lines) and 
0 (non-magnetic; dashed lines) clouds as a function of time. The 
thick (thin) lines are for the strong (weak) outflow cases with $f=0.1\ 
(0.01)$. Evidently, appreciable replenishment of turbulence requires 
both a strong outflow and a high rate of star formation, a condition 
that is met only with the combination $\Gamma_0=0$ and $f=0.1$. 
\label{fig:12}}
\end{figure}

\end{document}